\documentclass[11pt,letterpaper]{article}
\usepackage[margin=1in]{geometry}
\usepackage{amsmath,amssymb,amsthm,amsfonts}
\usepackage{mathtools}
\usepackage{enumitem}
\usepackage{booktabs}
\usepackage{multirow}
\usepackage[dvipsnames]{xcolor}
\definecolor{figureblue}{HTML}{1455C0}
\usepackage{tikz}
\usepackage{graphicx}
\usepackage{float}
\usepackage{caption}
\usepackage[numbers,square]{natbib}
\usepackage{microtype}
\usepackage[colorlinks=true,linkcolor=teal,citecolor=teal,urlcolor=teal]{hyperref}
\usetikzlibrary{arrows.meta}

\newtheorem{theorem}{Theorem}[section]
\newtheorem{lemma}[theorem]{Lemma}

\newcommand{\Greedy}{Greedy}
\newcommand{\MarkingGreedy}{Marking-Greedy}
\newcounter{algorithm}
\newcommand{\AlgorithmName}{Reserve-Retain-Recover}
\newcommand{\PhaseOneName}{Reserve}
\newcommand{\PhaseTwoName}{Retain}
\newcommand{\PhaseThreeName}{Recover}

\newcommand{\N}{N}
\newcommand{\R}{R}
\newcommand{\K}{K}
\newcommand{\Survival}{r}
\newcommand{\ALG}{\mathsf{ALG}}
\newcommand{\OPT}{\mathsf{OPT}}
\newcommand{\E}{\mathbb{E}}

\newenvironment{proofof}[1]
  {\begin{proof}[Proof of #1]}
  {\end{proof}}

\title{Improved Bounds for Online Matching with Random Edge Arrivals}
\author{
  Yilong Feng\thanks{State Key Lab of IOTSC, University of Macau.
    Email: \href{mailto:yc37459@um.edu.mo}{\texttt{yc37459@um.edu.mo}}.}
  \and
  Haolong Li\thanks{State Key Lab of IOTSC, University of Macau.
    Email: \href{mailto:yc47435@um.edu.mo}{\texttt{yc47435@um.edu.mo}}.}
  \and
  Zonghan Yang\thanks{Shanghai Jiao Tong University.
    Email: \href{mailto:fstqwq@sjtu.edu.cn}{\texttt{fstqwq@sjtu.edu.cn}}.
    Part of this work was done while the author was visiting the
    State Key Lab of IOTSC, University of Macau.}
}
\date{}

\begin{document}
\maketitle

\begin{abstract}
We study the unweighted online matching problem with random edge arrivals, where each edge arrives at an independent time drawn uniformly from $[0,1]$, and the algorithm must immediately and irrevocably accept or reject it while maintaining a matching.
Guruganesh and Singla (IPCO, 2017) first broke the $1/2$ barrier of the naive \Greedy{} algorithm by giving a $(1/2+\varepsilon)$-competitive algorithm for a small constant $\varepsilon>0$, a guarantee that has remained the best known for nearly a decade.
We give a simple three-phase algorithm that achieves a competitive ratio of $0.5445$ by deliberately rejecting selected early edges to reserve their endpoints for later matching opportunities.
With a different choice of parameters, the same algorithm is $0.5535$-competitive on triangle-free graphs, including bipartite graphs.
We also establish competitive ratio upper bounds below $0.6964$ for general graphs and $0.7043$ for bipartite graphs via exact rational evaluation of an optimal-policy recurrence on fixed graphs, even when the input graph is known up to isomorphism.
\end{abstract}
\newpage
\section{Introduction}
\label{sec:introduction}

Online matching has been a central topic in online algorithms since the seminal work of \citet*{conf/stoc/KarpVV90} on \emph{online bipartite matching}.
In this one-sided vertex-arrival model, the optimal competitive ratio is $1-1/e$ under adversarial arrivals~\cite{journals/sigact/BirnbaumM08,conf/soda/DevanurJK13,conf/sosa/EdenFFS21}, while strictly larger ratios are achievable under a uniformly random arrival order~\cite{conf/stoc/KarandeMT11,conf/stoc/MahdianY11}.
In the more general \emph{edge-arrival model}, edges of an underlying graph arrive sequentially, and the algorithm must immediately and irrevocably accept or reject each arriving edge while maintaining a matching.
Its performance is measured by the \emph{competitive ratio}, the worst-case ratio of its expected matching size to the maximum matching size.

Under adversarial edge arrivals, the simple \Greedy{} algorithm, which matches every arriving edge whenever possible, is $1/2$-competitive.
\citet*{conf/focs/GamlathKMSW19} proved that no randomized algorithm can do better even on bipartite graphs.
A natural question is whether \emph{random edge arrivals}, where the adversary chooses the graph while the edges arrive in a uniformly random order, permit a substantially better guarantee.
Unfortunately, \Greedy{}'s competitive ratio remains $1/2$, even on bipartite graphs~\cite{journals/rsa/DyerF91,conf/ipco/GuruganeshS17}.
\citet{conf/ipco/GuruganeshS17} first broke the $1/2$ barrier, achieving $0.50009$ on bipartite graphs and $1/2+\varepsilon$ on general graphs for a small constant $\varepsilon>0$, assuming the number of edges $m$ is known in advance.\footnote{Their \MarkingGreedy{} algorithm and analysis extend to independent uniform arrival times in $[0, 1]$ with the same competitive guarantees, using time thresholds for the phases without advance knowledge of $m$. We discuss the distinctions between these arrival models in Section~\ref{sec:related-work}.}
On the impossibility side, the best known upper bounds are $47/60\approx 0.7833$ for general graphs~\cite{FavrholdtVatshelleManuscript} and $23/28\approx 0.8214$ for bipartite graphs~\cite{conf/ipco/GuruganeshS17}.

\subsection{Our Model and Results}
\label{sec:model}

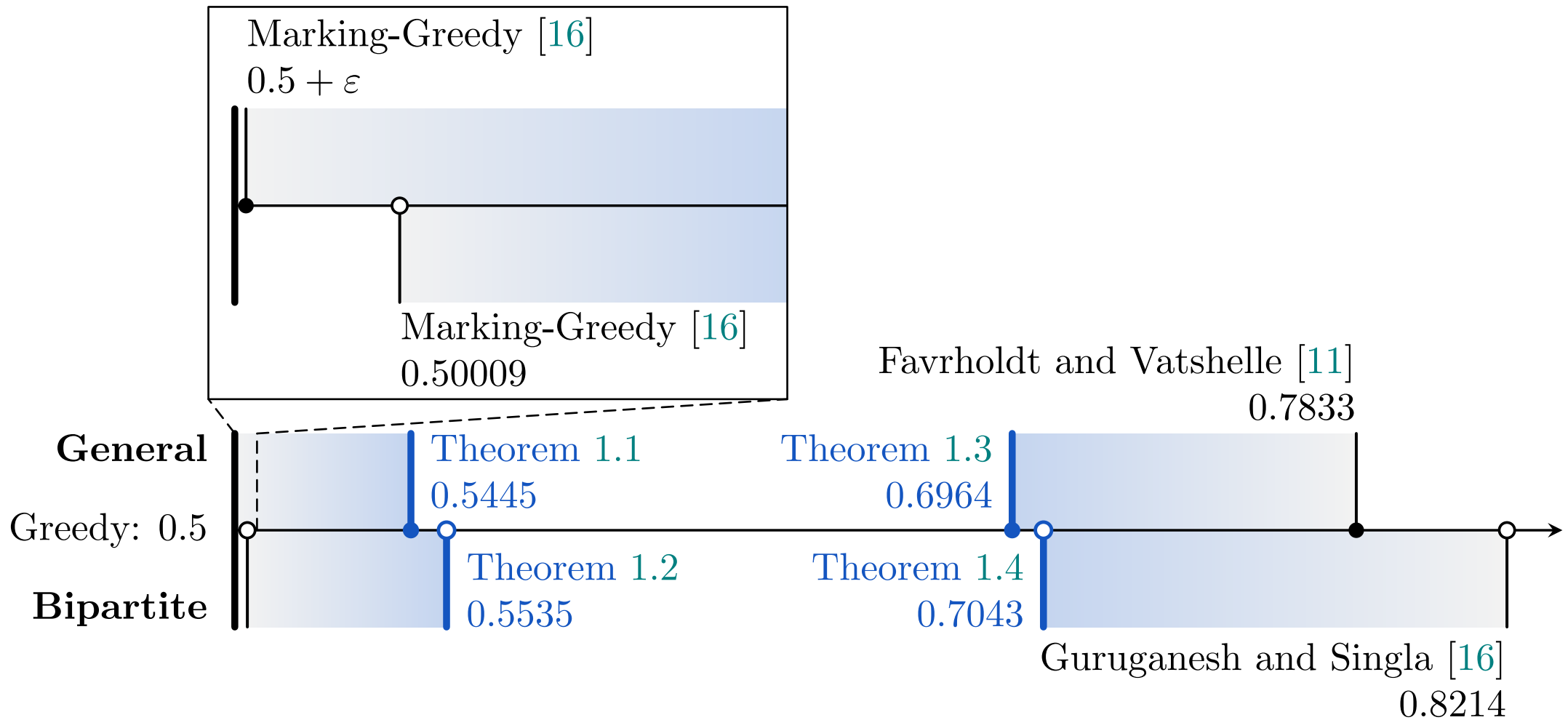
\begin{figure}[H]
\centering
\begin{tikzpicture}[x=1.5cm,y=0.8cm,font=\small,line cap=round,line join=round,>=stealth]

  \def\bandheight{1.10}
  \shade[left color=black!5,right color=figureblue!25] (0,0) rectangle (1.068,\bandheight);
  \shade[left color=figureblue!25,right color=black!5] (4.7136,0) rectangle (6.8,\bandheight);
  \shade[left color=figureblue!25,right color=black!5] (4.9032,-\bandheight) rectangle (7.7136,0);
  \shade[left color=black!5,right color=figureblue!25] (3.25pt,-\bandheight) rectangle (1.284,0);
  \draw[->,black,line width=.65pt] (0,0)--(8.05,0);
  \draw[black,line width=1.8pt] (0,-\bandheight)--(0,\bandheight);
  \node[anchor=east,align=center,inner sep=0pt] at (-.16,0) {\Greedy{}: $0.5$};
  \node[anchor=north east,inner sep=0pt] at (-.16,\bandheight) {\textbf{General}};
  \node[anchor=south east,inner sep=0pt] at (-.16,-\bandheight) {\textbf{Bipartite}};
  \draw[black,line width=.75pt] (3.25pt,0)--(3.25pt,-\bandheight);
  \filldraw[fill=white,draw=black,line width=.7pt] (3.25pt,0) circle (2pt);

  \foreach \x in {1.068,4.7136} {
    \draw[figureblue,line width=1.8pt] (\x,0)--(\x,\bandheight);
    \fill[figureblue] (\x,0) circle (2.1pt);
  }
  \node[anchor=north west,align=left,inner sep=0pt,text=figureblue] at (1.18,\bandheight)
    {Theorem~\ref{thm:main}\\$0.5445$};
  \node[anchor=north east,align=right,inner sep=0pt,text=figureblue] at (4.60,\bandheight)
    {Theorem~\ref{thm:upper-bound}\\$0.6964$};
  \draw[black,line width=.75pt] (6.8,0)--(6.8,\bandheight);
  \node[anchor=south east,align=right,inner sep=0pt,text=black] at (6.8,1.25)
    {\citet{FavrholdtVatshelleManuscript}\\$0.7833$};
  \fill[black] (6.8,0) circle (2pt);
  \draw[figureblue,line width=1.8pt] (1.284,0)--(1.284,-\bandheight);
  \filldraw[fill=white,draw=figureblue,line width=.9pt] (1.284,0) circle (2.1pt);
  \node[anchor=south west,align=left,inner sep=0pt,text=figureblue] at (1.40,-\bandheight)
    {Theorem~\ref{thm:main-tf}\\$0.5535$};
  \draw[black,line width=.75pt] (7.7136,0)--(7.7136,-\bandheight);
  \node[anchor=north east,align=right,inner sep=0pt,text=black] at (7.7136,-1.25)
    {\citet{conf/ipco/GuruganeshS17}\\$0.8214$};
  \filldraw[fill=white,draw=black,line width=.7pt] (7.7136,0) circle (2pt);
  \draw[figureblue,line width=1.8pt] (4.9032,0)--(4.9032,-\bandheight);
  \node[anchor=south east,align=right,inner sep=0pt,text=figureblue] at (4.79,-\bandheight)
    {Theorem~\ref{thm:bipartite-upper}\\$0.7043$};
  \filldraw[fill=white,draw=figureblue,line width=.9pt] (4.9032,0) circle (2.1pt);

  \draw[black,densely dashed,line width=.5pt] (0,\bandheight)--(-.16,1.49);
  \draw[black,densely dashed,line width=.5pt] (5.75pt,0)--(5.75pt,\bandheight)--(3.35,1.49);
  \begin{scope}[shift={(0,3.69)}]
    \begin{scope}
      \clip (-.16,-2.20) rectangle (3.35,2.26);
      \shade[left color=black!5,right color=figureblue!25] (2.9pt,0) rectangle (3.50,\bandheight);
      \shade[left color=black!5,right color=figureblue!25] (1,-\bandheight) rectangle (3.50,0);
      \draw[black,line width=.65pt] (0,0)--(3.50,0);
      \draw[black,line width=1.8pt] (0,-\bandheight)--(0,\bandheight);
      \draw[black,line width=.75pt] (2.9pt,0)--(2.9pt,\bandheight);
      \fill[black] (2.9pt,0) circle (2pt);
      \node[anchor=south west,align=left,inner sep=0pt,text=black] at (2.9pt,1.12)
        {\MarkingGreedy{}~\cite{conf/ipco/GuruganeshS17}\strut\\$0.5+\varepsilon$\strut};
      \node[anchor=north west,align=left,inner sep=0pt,text=black] (bipartite) at (1,-1.18)
        {\MarkingGreedy{}~\cite{conf/ipco/GuruganeshS17}\\$0.50009$};
      \draw[black,line width=.75pt] (1,0)--(1,-\bandheight);
      \filldraw[fill=white,draw=black,line width=.7pt] (1,0) circle (2pt);
    \end{scope}
    \draw[black,line width=.5pt] (-.16,-2.20) rectangle (3.35,2.26);
  \end{scope}
\end{tikzpicture}
\caption{Competitive ratio overview. Our results are marked in \textcolor{figureblue}{blue}.}
\label{fig:competitive-ratios}
\end{figure}

We study online matching with random edge arrivals in continuous time.
For an unknown finite simple graph $G(V,E)$ with a maximum matching $M^*$ of size $\OPT=|M^*|$, each edge $e\in E$ arrives at an independent time $t_e\sim U[0,1]$.
The algorithm observes each arriving edge together with its arrival time, but knows neither $G$ nor $m=|E|$.
Let $\ALG$ be the size of the matching it produces; its competitive ratio is defined as $\inf_G\{\E[\ALG]/\OPT\}$.
We improve both the lower and upper bounds on the optimal competitive ratio, as shown in Figure~\ref{fig:competitive-ratios}.

\paragraph{Lower bounds.}
We propose a three-phase algorithm called \emph{\AlgorithmName{}}, which partitions the edge stream by the observed arrival times without prior knowledge of $G$ or $m$.
We analyze its competitive ratio in  Sections~\ref{sec:global}, \ref{sec:poly} and  \ref{sec:local}.
With a different choice of parameters, the same algorithm achieves a stronger guarantee on triangle-free graphs; see Appendix~\ref{app:triangle-free}.

\begin{theorem}[General lower bound]
\label{thm:main}
There exists a $0.5445$-competitive algorithm for online matching with random edge arrivals.
\end{theorem}

\begin{theorem}[Triangle-free lower bound]
\label{thm:main-tf}
There exists a $0.5535$-competitive algorithm for online matching with random edge arrivals on triangle-free graphs, including bipartite graphs.
\end{theorem}

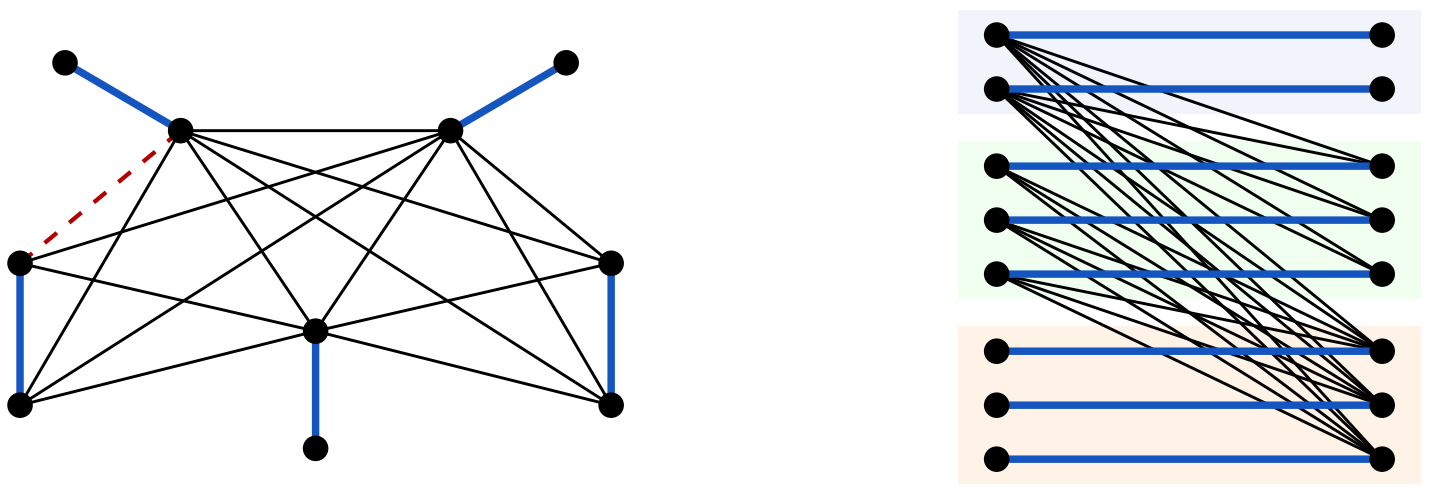
\begin{figure}[H]
\centering
\begin{tikzpicture}[scale=.85,yscale=1.2,
  vertex/.style={circle,fill=black,inner sep=1.7pt},
  ordinary/.style={draw=black,line width=.55pt},
  matching/.style={draw=figureblue,line width=1.4pt},
  missing/.style={red!70!black,dashed,line width=.8pt},
  every label/.style={font=\footnotesize,inner sep=1.5pt}]
\begin{scope}
  \node[align=center] at (0,2.15) {General: $G_{10}$\\$0.696327$};
  \coordinate (7) at (-1.05,.78);
  \coordinate (8) at (1.05,.78);
  \coordinate (9) at (0,-.52);
  \coordinate (2) at (-1.95,1.22);
  \coordinate (3) at (1.95,1.22);
  \coordinate (4) at (0,-1.28);
  \coordinate (6) at (-2.3,-.08);
  \coordinate (1) at (-2.3,-1);
  \coordinate (0) at (2.3,-.08);
  \coordinate (5) at (2.3,-1);
  \draw[ordinary] (7)--(8)--(9)--(7);
  \foreach \v in {8,9} \draw[ordinary] (6)--(\v);
  \foreach \u in {1,0,5}
    \foreach \v in {7,8,9} \draw[ordinary] (\u)--(\v);
  \draw[missing] (6)--(7);
  \draw[matching] (0)--(5) (1)--(6) (2)--(7) (3)--(8) (4)--(9);
  \foreach \v in {7,8} \node[vertex] at (\v) {};
  \foreach \v in {1,2,6} \node[vertex] at (\v) {};
  \foreach \v in {0,3,5} \node[vertex] at (\v) {};
  \node[vertex] at (4) {};
  \node[vertex] at (9) {};
\end{scope}
\begin{scope}[xshift=6.8cm]
  \node[align=center] at (0,2.15) {Bipartite: $B_{2,3,3}$\\$0.704284$};
  \fill[figureblue!6] (-1.8,1.56) rectangle (1.8,0.89);
  \fill[green!6] (-1.8,0.71) rectangle (1.8,-0.31);
  \fill[orange!10] (-1.8,-0.49) rectangle (1.8,-1.51);
  \coordinate (L0) at (-1.5,1.4); \coordinate (R0) at (1.5,1.4);
  \coordinate (L1) at (-1.5,1.05); \coordinate (R1) at (1.5,1.05);
  \coordinate (L2) at (-1.5,0.55); \coordinate (R2) at (1.5,0.55);
  \coordinate (L3) at (-1.5,0.2); \coordinate (R3) at (1.5,0.2);
  \coordinate (L4) at (-1.5,-0.15); \coordinate (R4) at (1.5,-0.15);
  \coordinate (L5) at (-1.5,-0.65); \coordinate (R5) at (1.5,-0.65);
  \coordinate (L6) at (-1.5,-1); \coordinate (R6) at (1.5,-1);
  \coordinate (L7) at (-1.5,-1.35); \coordinate (R7) at (1.5,-1.35);
  \draw[ordinary] (L0)--(R2);
  \draw[ordinary] (L0)--(R3);
  \draw[ordinary] (L0)--(R4);
  \draw[ordinary] (L0)--(R5);
  \draw[ordinary] (L0)--(R6);
  \draw[ordinary] (L0)--(R7);
  \draw[ordinary] (L1)--(R2);
  \draw[ordinary] (L1)--(R3);
  \draw[ordinary] (L1)--(R4);
  \draw[ordinary] (L1)--(R5);
  \draw[ordinary] (L1)--(R6);
  \draw[ordinary] (L1)--(R7);
  \draw[ordinary] (L2)--(R5);
  \draw[ordinary] (L2)--(R6);
  \draw[ordinary] (L2)--(R7);
  \draw[ordinary] (L3)--(R5);
  \draw[ordinary] (L3)--(R6);
  \draw[ordinary] (L3)--(R7);
  \draw[ordinary] (L4)--(R5);
  \draw[ordinary] (L4)--(R6);
  \draw[ordinary] (L4)--(R7);
  \draw[matching] (L0)--(R0);
  \draw[matching] (L1)--(R1);
  \draw[matching] (L2)--(R2);
  \draw[matching] (L3)--(R3);
  \draw[matching] (L4)--(R4);
  \draw[matching] (L5)--(R5);
  \draw[matching] (L6)--(R6);
  \draw[matching] (L7)--(R7);
  \foreach \i in {0,...,7} {\node[vertex] at (L\i) {}; \node[vertex] at (R\i) {};}
\end{scope}
\end{tikzpicture}
\caption{Hard instances for the upper bounds in Theorems~\ref{thm:upper-bound} and~\ref{thm:bipartite-upper}.}
\label{fig:upper-construction}
\end{figure}

\paragraph{Upper bounds.}
In Section~\ref{sec:upper-proof},
we give the input graph to the algorithm up to isomorphism, while choosing its vertex labeling and edge arrival order uniformly at random.
The optimal online algorithm under this uncertainty can be described as a dynamic program.
Evaluating the recurrence on the hard instances $G_{10}$ and $B_{2,3,3}$ (Figure~\ref{fig:upper-construction}) yields the following upper bounds.

\begin{theorem}[General upper bound]\label{thm:upper-bound}
Even when the input graph is known up to isomorphism, no online algorithm has a competitive ratio greater than $0.6964$ on general graphs.
\end{theorem}

\begin{theorem}[Bipartite upper bound]\label{thm:bipartite-upper}
Even when the input graph is known up to isomorphism, no online algorithm has a competitive ratio greater than $0.7043$ on bipartite graphs.
\end{theorem}

\subsection{Our Algorithm and Techniques}
\label{sec:algorithm-techniques}

We now present \AlgorithmName{} (Algorithm~\ref{alg:reserve-retain-recover}) and the ideas behind the proof of Theorem~\ref{thm:main}.

\vspace{6pt}
\noindent\makebox[\textwidth][c]{\fbox{
\begin{minipage}{0.96\textwidth}
\vspace{3pt}
\refstepcounter{algorithm}\label{alg:reserve-retain-recover}\textbf{Algorithm \thealgorithm: \AlgorithmName{}} \hfill (Parameters: $0\le a<b\le 1$ and $0\le p\le 1$)
\vspace{3pt}
\hrule
\vspace{5pt}
\textbf{Initialize:} matching $M \leftarrow \varnothing$, shadow matching $T \leftarrow \varnothing$, reserved vertices $X \leftarrow \varnothing$.
\begin{itemize}[leftmargin=1.6em, itemsep=1pt, topsep=2pt]
    \item \textbf{Phase 1 (\PhaseOneName{}, $t\in [0,a)$):} For each arriving edge $e=(u,v)$, if both $u,v$ are unmatched in $T$, add $e$ to $T$ (calling it a \emph{shadow edge}) and flip an independent coin: w.p. $p$, discard $e$ and add both $u,v$ to $X$; w.p. $1-p$, add $e$ to $M$. Otherwise, discard $e$.
    \item \textbf{Phase 2 (\PhaseTwoName{}, $t\in [a,b)$):} For each arriving edge $e=(u,v)$, if both $u,v$ are unmatched in $T$, add $e$ to both $T$ and $M$ (also calling it a \emph{shadow edge}). Otherwise, discard $e$.
    \item \textbf{Phase 3 (\PhaseThreeName{}, $t\in [b,1]$):} Let $Y$ be the set of vertices left unmatched in $T$ at time $b$. For each arriving edge $e=(u,v)$, if both $u,v$ are unmatched in $M$ and not both in $X$ (i.e., $e$ is an $XY$-edge or a $YY$-edge), add $e$ to $M$. Otherwise, discard $e$.
\end{itemize}
\vspace{2pt}
\end{minipage}
}}

\paragraph{Algorithm design.}

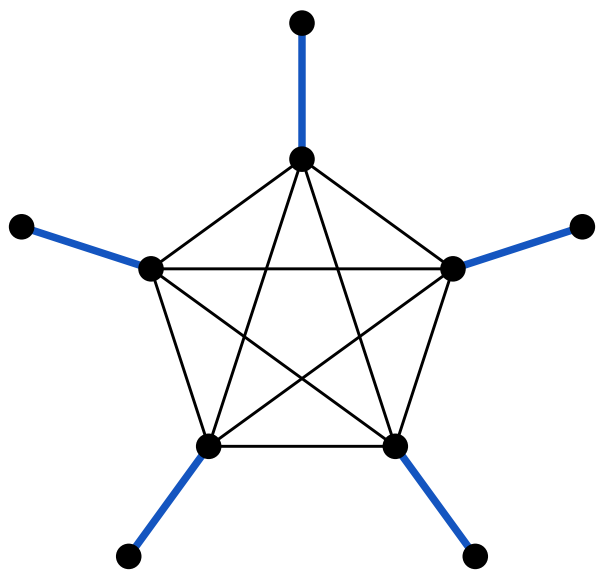
\begin{figure}[h]
\centering
\begin{tikzpicture}[
  vertex/.style={circle,fill=black,inner sep=1.7pt},
  ordinary/.style={draw=black,line width=.55pt},
  matching/.style={draw=figureblue,line width=1.4pt},
  font=\small
]
  \foreach \i/\angle in {1/90,2/18,3/-54,4/-126,5/-198} {
    \coordinate (x\i) at (\angle:1.05);
    \coordinate (y\i) at (\angle:1.95);
  }
  \foreach \i in {1,...,4} {
    \pgfmathtruncatemacro{\next}{\i+1}
    \foreach \j in {\next,...,5} {
      \draw[ordinary] (x\i)--(x\j);
    }
  }
  \foreach \i in {1,...,5} {
    \draw[matching] (x\i)--(y\i);
  }
  \foreach \i in {1,...,5} {
    \node[vertex] at (x\i) {};
    \node[vertex] at (y\i) {};
  }
\end{tikzpicture}
\caption{A sunflower with a $K_5$ core. The \protect\tikz[baseline=(X.base)]{\protect\node[inner sep=0pt,text=figureblue](X){blue};\protect\draw[figureblue,line width=1.2pt]([yshift=-2.3pt]X.base west)--([yshift=-2.3pt]X.base east);} leaf edges form the unique maximum matching.}
\label{fig:sunflower}
\end{figure}

Consider a \textit{sunflower} consisting of $K_n$ with one private leaf attached to each core vertex, as shown in Figure~\ref{fig:sunflower}.
The unique maximum matching contains all $n$ leaf edges, whereas \Greedy{} matches only $n/2+\Theta(\log n)$ edges in expectation under random edge arrivals, giving a ratio of $1/2+\Theta(\log n/n)$ on this family.

In this graph, the $\binom{n}{2}$ core edges vastly outnumber the $n$ leaf edges: within the first $\Theta(n\log n)$ arrivals, \Greedy{} already matches almost all core vertices, mostly via core edges that each block two edges of the unique perfect matching.
Rejecting all edges selected by \Greedy{} in this early prefix to reserve their endpoints, and later matching only reserved vertices to remaining unmatched vertices, recovers almost all leaf edges and achieves a ratio of $1-O(1/\log n)$.
On general graphs, this early-saturation phenomenon is captured by the \emph{Hastiness Lemma} used by \citet{conf/ipco/GuruganeshS17}, which shows that whenever \Greedy{} performs close to $1/2$, it accepts most of its edges early.
Based on this property, their \MarkingGreedy{} algorithm maintains the shadow greedy matching $T$ up to time $b$, randomly rejecting a $p$-fraction of shadow edges to form a reserved set $X$, and in $[b,1]$ seeks additional matches between $X$ and the set $Y$ of vertices left unmatched in $T$.

On graphs where \Greedy{} already performs well, however, rejecting early shadow edges may sacrifice edges of $M^*$ without creating compensating recovery opportunities.
In fact, the expected rate at which \Greedy{} accepts edges decreases over time, and in hard instances is heavily concentrated near the very beginning of the stream.
This allows \AlgorithmName{} to subdivide $[0,b)$ more finely into a reserve phase $[0,a)$ and an intermediate retain phase $[a,b)$: we only reject shadow edges in the earliest phase $[0,a)$, accept every new shadow edge during $[a,b)$, and defer recovery of the reserved vertices $X$ to the final phase $[b,1]$.
At time $b$, the vertices unmatched in $M$ partition into $X$ and $Y$, and in the recovery phase $[b,1]$ the algorithm greedily accepts both $XY$-edges and $YY$-edges.
Including $YY$-edges provides additional matches, though they may also compete with $XY$-edges for endpoints in $Y$; our analysis accounts for both together.

\paragraph{Conditional arrival times.}
To analyze the recovery phase, we condition on the shadow matching $T$ at time $b$ and the arrival times of its edges, which fixes the unmatched vertex set $Y$.
Consider a shadow edge $(x_1,x_2)$ selected at time $s<a$ and an edge $f$ between $\{x_1,x_2\}$ and $Y$.
Had $f$ arrived before $s$, both of its endpoints would have been unmatched in $T$ and $T$ would have selected $f$; after time $s$, $f$ no longer affects $T$.
Hence the conditioning only reveals that $f$ arrives after $s$, so its arrival time is uniform on $(s,1]$, independently of the other edges, and $f$ arrives in the recovery phase $[b,1]$ w.p.
\begin{equation*}
    q_s=\frac{1-b}{1-s}.
\end{equation*}
Similarly, every $YY$-edge arrives after time $b$.
Once the reservation choices are fixed, the recovery phase is \Greedy{} on the $XY$- and $YY$-edges that arrive in $[b,1]$: each $XY$-edge appears independently w.p. $q_s$, every $YY$-edge appears w.p. $1$, and these edges arrive in a uniform random order.

\paragraph{Coupling two greedy matchings.}
Fix a maximum matching $M^*$ and a shadow edge $e=(x_1,x_2)$ from the reserve phase.
Reserving $x_1,x_2$ gives up $e$, so we must account for what the recovery phase gains in return: besides $x_1$ and $x_2$, we track their $k\le 2$ partners in $Y$ under $M^*$, since an edge from $x_i$ to its $Y$-partner can be blocked during recovery only if at least one of its endpoints is already matched.
To measure how the coin flip for $e$ affects these vertices, we fix the reservation choices of all other shadow edges and couple the two recovery runs of \Greedy{}, without $x_1,x_2$ (w.p. $1-p$, when $e$ is accepted) and with $x_1,x_2$ in $X$ (w.p. $p$, when $x_1,x_2$ are reserved), using the same arrival order on shared edges.
Let $\N$ and $\R$ count the matched vertices among $\{x_1,x_2\}$ and among their $Y$-partners in the run with $x_1,x_2$, and let $\K$ count the matched $Y$-partners in the run without them.
Averaging over the coin flip for $e$, recovery matches $p\,\E[\N]$ vertices in $\{x_1,x_2\}$ and $p\,\E[\R]+(1-p)\,\E[\K]$ of their $Y$-partners in expectation.
Weighting these counts by $3/2$ and $1$ and using $1-p\ge p/2$ (for $p\le 2/3$) reduces our local recovery bound (Lemma~\ref{lem:early-pair-recovery}),
\begin{equation*}
    \frac32 p\,\E[\N]+p\,\E[\R]+(1-p)\,\E[\K]\ge \frac{p}{2}\E[3\N+2\R+\K]\ge \frac{3p}{2}k,
\end{equation*}
to the coupled comparison $\E[3\N+2\R+\K]\ge 3k$, proven in Section~\ref{sec:local} (Lemma~\ref{lem:greedy-comparison}).

\paragraph{The global bound.}
Summing the local bounds over the shadow edges from the reserve phase still leaves graph-dependent terms, such as the number of shadow edges selected in each phase and the number of edges in $M^*$ whose endpoints are both unmatched in $T$ at time $b$.
We split the probability that both endpoints of $e\in M^*$ are unmatched in $T$ at time $t$ into
\begin{equation*}
    \Pr[e \text{ arrives after } t]\cdot \Pr[\text{both endpoints of } e \text{ are unmatched at } t \text{ in } G\setminus\{e\}].
\end{equation*}
Let $\Survival_e(t)$ denote the latter probability, which we call the \emph{survival probability} of $e$ up to time $t$.
In Section~\ref{sec:global}, we express all graph-dependent terms as a global lower bound that is linear in $\Survival(t)$, the average of $\Survival_e(t)$ over $e\in M^*$, retaining information at both phase boundaries and throughout the retain phase rather than relying only on a hastiness bound for early edges.
A key observation is that $\Survival_e(t)$ (and hence $\Survival(t)$) is a convex combination of the powers $(1-t)^{n}$ for integers $n\ge 0$.
Intuitively, $(1-t)^{n}$ is the probability that none of $n$ independent edges have arrived by time $t$, which contributes to the event that both endpoints of $e$ are unmatched at time $t$.
We formalize this observation in Lemma~\ref{lem:poly-expansion}.
By linearity, it suffices to evaluate the global bound on each power separately.
We then choose $a,b,p$ and verify the resulting scalar inequalities to obtain the competitive ratio of $0.5445$.

\subsection{\texorpdfstring{{Related Work}}{Related Work}}
\label{sec:related-work}

Due to the vast literature on online matching, here we only review the most relevant ones.
For broader accounts, we refer to the excellent surveys of \citet{journals/fttcs/Mehta13} and \citet*{journals/sigecom/HuangTW24}.

\paragraph{Online bipartite matching with random vertex arrivals.}
\citet{conf/soda/GoelM08} showed that \Greedy{} is $(1-1/e)$-competitive in this model.
\citet*{conf/stoc/KarandeMT11} and \citet{conf/stoc/MahdianY11} analyzed Ranking, obtaining ratios of $0.653$ and $0.696$, respectively.
\citet*{journals/talg/HuangTWZ19} extended Ranking to the vertex-weighted case, where offline vertices have nonnegative weights, and achieved a competitive ratio of $0.6534$.
The ratio was later improved to $0.6629$ by \citet{conf/wine/JinW21}, $0.6862$ by \citet{conf/sigecom/PengT25}, and $0.6980$ by \citet{journals/corr/abs-2608-12176}.

\paragraph{Variants of online matching with edge arrivals.}
Competitive ratios better than $0.5$ are known under structural or distributional relaxations for online matching with edge arrivals.
\citet*{journals/algorithmica/BuchbinderST19} gave a $5/9$-competitive algorithm on trees with adversarial edge arrivals; with free disposal, \citet{conf/wine/JiangZ24} achieved $5/8$ on forests and $2/3$ on growing trees, where each arriving edge introduces a new leaf.
\citet{journals/rsa/DyerF91} gave a nearly tight ratio of $0.7690$ for \Greedy{} on trees under random edge arrivals.
\citet*{conf/icalp/GravinTW21} assumed that edges of a given bipartite graph arrive in an order chosen by an oblivious adversary and are independently realized with known probabilities, and gave a $0.503$-competitive algorithm.
When edges are sampled independently and identically from a known distribution, \citet*{journals/corr/abs-2606-16537} beat $1-1/e$ under integral arrival rates.
For edge-weighted online matching with random edge arrivals, \citet*{journals/mor/EzraFGT25} gave a $1/4$-competitive algorithm.

\paragraph{Arrival times and advance information.}
Assuming that elements arrive at independent uniform times in $[0,1]$ is a common convention for random order models in the online algorithms literature, e.g., in online bipartite matching with random vertex arrivals~\cite{journals/talg/HuangTWZ19,conf/wine/JinW21,conf/sigecom/PengT25,journals/corr/abs-2608-12176}, prophet secretary~\cite{conf/soda/Chen0LT25,journals/mp/CorreaSZ21,journals/siamcomp/EhsaniHKS24}, and sample-driven optimal stopping~\cite{journals/mor/0001C0S24}.
For random edge arrivals, the algorithms of \citet{conf/ipco/GuruganeshS17} assume that $m$ is known in advance; when $m$ is unknown and the algorithm observes only the arrival order, they proved an upper bound of $2/3$ by hiding which of several graphs of different sizes is presented.
Together with our observed-time model and the setting where the algorithm knows the input graph up to isomorphism, these information settings satisfy
\[
 \text{unknown }m\ \preceq\ \text{observed times}\ \preceq\ \text{known }m\ \preceq\ \text{known graph},
\]
where $A\preceq B$ means that every randomized strategy in $A$ can be simulated in $B$ with the same expected matching size on each graph. For example:
\begin{itemize}
\item To simulate observed times in the known-$m$ permutation model, sample $m$ independent uniform timestamps in $[0,1]$, and assign the $i$-th smallest timestamp to the $i$-th arriving edge; this reproduces the desired joint distribution of edge order and arrival times.
\item Conversely, when $m$ is also given in the arrival-time model, using the edge order alone recovers the known-$m$ permutation model.
\end{itemize}
Thus, when $m$ is known, the arrival-time and random-permutation models are equivalent.
Hence our lower bound also holds in the known-$m$ permutation model, and our upper bounds hold in all four settings.

\section{From Local Recovery to a Global Lower Bound}
\label{sec:global}

In this section, we give a lower bound on the competitive ratio of \AlgorithmName{}.
For simplicity of analysis, we assume parameters $0\le a<b\le 3/8$ and $0\le p\le 2/3$.

\paragraph{Notation.}
For any $F\subseteq E$, let $V(F)$ be the set of vertices covered by $F$.
For a graph $J$, write $J-e$, $J-z$, and $J-V(e)$ for the subgraphs obtained by deleting edge $e$, vertex $z$ (and its incident edges), and both endpoints of $e$, respectively.
For disjoint vertex sets $A$ and $B$, we call an edge $(u,v)$ with $u\in A$ and $v\in B$ an \emph{$AB$-edge}.

\smallskip

Let $T_t$ be the shadow matching at time $t\in [0,b]$, and partition $V$ into $V_1=V(T_a)$, $V_2=V(T_b)\setminus V_1$, and $Y=V \setminus V(T_b)$.
By definition, the algorithm matches all edges in $T_b\setminus T_a$, while each edge in $T_a$ gets matched w.p. $1-p$.
Let $Z_{XY}$ and $Z_{YY}$ be the numbers of $XY$-edges and $YY$-edges matched in Phase 3, respectively.
Let $H_b$ record the shadow matching $T_b$ and the arrival times of its edges.\footnote{The remaining randomness consists of the arrival times of edges outside $T_b$, and the random reservations.}
We have
\begin{equation}
\label{eq:prefix-suffix-decomposition}
    \E[\ALG\mid H_b] = |T_b| - p\cdot |T_a|
    +\E[Z_{XY}+Z_{YY}\mid H_b].
\end{equation}
We first lower-bound the contribution from Phase 3.

\subsection{A Local Recovery Bound}
\label{sec:pair-recovery}

Consider a shadow edge $e=(x_1,x_2)\in T_a$.
An edge of $M^*$ may connect an endpoint of $e$ to a vertex $v\in Y$.
We call such a vertex $v$ a reference vertex of $e$.
Since $M^*$ is a matching, $e$ has at most two reference vertices.
Let their number be $k_e\in \{0,1,2\}$.

We track two quantities during Phase 3.
Let $\hat \N_e$ (resp. $\hat \R_e$) be the number of vertices in $\{x_1,x_2\}$ (resp. reference vertices of $e$) matched during Phase 3.
A reference vertex contributes to $\hat \R_e$ whenever it is matched, regardless of which edge matches it.
In particular, if $e$ is not reserved, then $\hat \N_e=0$, but its reference vertices may still get matched during Phase 3.

The following lemma lower-bounds (the weighted sum of) the expected contribution from $\hat \N_e$ and $\hat \R_e$ in terms of $k_e$.

\begin{lemma}
\label{lem:early-pair-recovery}
Conditional on $H_b$, for any $p\in [0,2/3]$ and every shadow edge $e\in T_a$, we have
\begin{equation*}
    \frac{3}{2}\cdot \E[\hat \N_e\mid H_b] + \E[\hat \R_e\mid H_b] \ge \frac{3p}{2}\cdot k_e.
\end{equation*}
\end{lemma}

Let $J$ be the graph on $X\cup Y$ containing all $YY$-edges and all $XY$-edges arriving after time $b$.
Let $B=J-V(e)$, and let $B_e$ be the graph obtained from $B$ by adding $x_1,x_2$ and all edges between $\{x_1,x_2\}$ and $Y$
that arrive after $b$.
We compare the two reservation choices for $e$, keeping all edge arrival times and the reservation choices for all other shadow edges the same.
Since $B$ excludes $x_1,x_2$, it is unchanged by the reservation choice for $e$.
Phase 3 runs \Greedy{} on $B_e$ if $e$ is reserved, and on $B$ otherwise.

Let $M(B)$ and $M(B_e)$ be the greedy matchings on these graphs, using the same edge arrival times.
Let $\N$ count the vertices in $\{x_1,x_2\}$ matched in $M(B_e)$, and let $\R$ and $\K$ count the reference vertices of $e$ matched in $M(B_e)$ and $M(B)$, respectively.
The following lemma implies Lemma~\ref{lem:early-pair-recovery}.
For continuity of presentation, we defer its proof to Section~\ref{sec:local}.

\begin{lemma}
\label{lem:greedy-comparison}
For the coupled greedy matchings $M(B)$ and $M(B_e)$, we have
\begin{equation*}
    \E[3\N+2\R+\K\mid H_b,B]\ge 3k_e.
\end{equation*}
\end{lemma}

Taking expectations over $B$, we obtain $\E[3\N+2\R+\K\mid H_b]\ge 3k_e$.
If $e$ is reserved, then $\hat \N_e=\N$ and $\hat \R_e=\R$; otherwise, $\hat \N_e=0$ and $\hat \R_e=\K$.
Since the random reservation of $e$ is independent of $\N,\R,\K$ conditional on $H_b$, we have
\begin{equation*}
    \frac{3}{2}\cdot \E[\hat \N_e\mid H_b] + \E[\hat \R_e\mid H_b]
    = \frac{p}{2}\cdot \E[3\N+2\R+\K\mid H_b] + \left(1-\frac{3p}{2}\right)\cdot \E[\K\mid H_b]
    \ge \frac{3p}{2}\cdot k_e,
\end{equation*}
where the inequality uses $p\le 2/3$ and $\K\ge 0$.
This proves Lemma~\ref{lem:early-pair-recovery}.

We next sum the lower bound in Lemma~\ref{lem:early-pair-recovery} over all shadow edges in $T_a$ and relate the resulting vertex counts to the number of edges accepted in Phase 3.

\subsection{Aggregating the Local Bounds}
\label{sec:suffix-agg}

We continue to fix $T_b$ and the arrival times of its edges.
Define
\begin{equation*}
    M^*_{1Y} = \{(u,v)\in M^*: u\in V_1, v\in Y\}, \qquad
    M^*_{YY} = \{(u,v)\in M^*: u,v\in Y\}.
\end{equation*}
Notice that each edge in $M^*_{1Y}$ connects an endpoint of exactly one shadow edge $e\in T_a$ to a reference vertex of $e$.
Moreover, since $M^*$ is a matching, different shadow edges have disjoint sets of reference vertices.
Consequently, we have
\begin{equation*}
    \sum_{e\in T_a} k_e = |M^*_{1Y}|.
\end{equation*}

We now use Lemma~\ref{lem:early-pair-recovery} to lower-bound the number of edges matched during Phase 3.

\begin{lemma}
\label{prop:suffix-bound}
Conditional on $H_b$, for any $p\in[0,2/3]$, we have
\begin{equation*}
    \E[Z_{XY} + Z_{YY}\mid H_b] \ge \frac{3p}{5}\cdot |M^*_{1Y}| + \frac{2}{5}\cdot |M^*_{YY}|.
\end{equation*}
\end{lemma}
\begin{proof}
Let
\begin{equation*}
    \R_{1Y} = \sum_{e\in T_a} \hat \R_e
\end{equation*}
be the total number of reference vertices matched during Phase 3.
Each matched $XY$-edge matches exactly one endpoint of a shadow edge $e\in T_a$.
Thus,
\begin{equation*}
     \sum_{e\in T_a} \hat \N_e = Z_{XY}.
\end{equation*}
By Lemma~\ref{lem:early-pair-recovery} and summing over all $e\in T_a$, we have
\begin{equation}
\label{eq:sum-pairs}
    \frac{3}{2}\cdot \E[Z_{XY}\mid H_b] + \E[\R_{1Y}\mid H_b] \ge \frac{3p}{2}\cdot |M^*_{1Y}|.
\end{equation}

Next, consider an edge $(u,v)\in M^*_{YY}$.
By definition of $Y$, $(u,v)$ arrives after time $b$.
When $(u,v)$ arrives, either the algorithm matches it, or at least one of $u,v$ is already matched.
Choose one matched endpoint for each edge in $M^*_{YY}$.
These chosen vertices are distinct and disjoint from the reference vertices counted by $\R_{1Y}$, since $M^*_{1Y}\cup M^*_{YY}$ is a matching.
Notice that Phase 3 matches exactly $Z_{XY} + 2\cdot Z_{YY}$ vertices in $Y$.
We therefore have
\begin{equation*}
    \R_{1Y} + |M^*_{YY}| \le Z_{XY} + 2\cdot Z_{YY}.
\end{equation*}

Taking expectations of the above inequality, together with Inequality~\eqref{eq:sum-pairs}, we have
\begin{align*}
    \frac{3p}{2}\cdot |M^*_{1Y}|+|M^*_{YY}|
    &\le \frac{3}{2}\cdot \E[Z_{XY}\mid H_b] + \E[\R_{1Y}\mid H_b]+|M^*_{YY}| \\
    &\le \frac{5}{2}\cdot \E[Z_{XY}\mid H_b] + 2\cdot \E[Z_{YY}\mid H_b]
    \le \frac{5}{2}\cdot \E[Z_{XY}+Z_{YY}\mid H_b].
\end{align*}
Dividing by $5/2$ proves the lemma.
\end{proof}

\subsection{\texorpdfstring{{Accounting for the \PhaseOneName{} and \PhaseTwoName{} Phases}}{Accounting for the Reserve and Retain Phases}}
\label{sec:combining}

Combining Equation~\eqref{eq:prefix-suffix-decomposition} with Lemma~\ref{prop:suffix-bound}, we have
\begin{equation*}
    \E[\ALG\mid H_b] \ge |T_b| - p\cdot |T_a| + \frac{3p}{5}\cdot |M^*_{1Y}| + \frac{2}{5}\cdot |M^*_{YY}|.
\end{equation*}
We next relate the RHS to $\OPT$ using two counts of vertices matched in $M^*$.

For any $t\in [0,b]$, let $U_t$ be the number of edges in $M^*$ with both endpoints unmatched in $T_t$.
In particular, we have $U_b=|M^*_{YY}|$.
Let $M^*_2=M^*\cap (T_b\setminus T_a)$ denote the edges in $M^*$ selected by the shadow matching during Phase 2.

\begin{lemma}
\label{lem:count1}
We have
\begin{equation*}
    |M^*_{1Y}| + |M^*_{YY}| \ge 2\cdot \OPT - 2\cdot |T_b| - U_a + |M^*_2|.
\end{equation*}
\end{lemma}
\begin{proof}
Since $T_b$ matches at most $2\cdot |T_b|$ vertices in $V(M^*)$, we have $|Y\cap V(M^*)|\ge 2\cdot \OPT - 2\cdot |T_b|$.

We next upper-bound $|Y\cap V(M^*)|$ by counting the contributions from different edge sets.
Each $V_1Y$-edge in $M^*$ contributes one vertex to $Y\cap V(M^*)$ and is counted by $|M^*_{1Y}|$.
Each $V_2Y$-edge in $M^*$ contributes one such vertex and is counted by $U_a$.
Each $YY$-edge in $M^*$ contributes two and is counted once by $U_a$ and once by $|M^*_{YY}|$.
Finally, every edge in $M^*_2$ is counted by $U_a$ but has both endpoints in $V_2$.
Therefore, $|Y\cap V(M^*)| \le |M^*_{1Y}| + |M^*_{YY}| + U_a - |M^*_2|$.

Combining the two inequalities proves the lemma.
\end{proof}

\begin{lemma}
\label{lem:count2}
We have
\begin{equation*}
    2(|T_b|-|T_a|) \ge U_a - U_b + |M^*_2|.
\end{equation*}
\end{lemma}
\begin{proof}
Notice that shadow edges selected during Phase 2 match exactly $2(|T_b|-|T_a|)$ vertices.
By definition, $U_a-U_b$ counts the edges in $M^*$ with both endpoints unmatched in $T_a$ and at least one endpoint matched in $T_b$.
Every edge in $M^*_2$ belongs to these $U_a-U_b$ edges and has both endpoints matched in $T_b$.
Since $M^*$ is a matching, these contributions count distinct vertices of $V_2$, and the lemma follows.
\end{proof}

\paragraph{Choice of the reservation probability.}
Recall that the expected number of edges matched before time $b$ is $|T_b| - p\cdot |T_a|$.
Combining this with Lemma~\ref{prop:suffix-bound} and Lemma~\ref{lem:count1}, the terms involving $|T_a|$ and $|T_b|$ can be written as
\begin{equation*}
    \left(1-\frac{6p}{5}\right)\cdot |T_b| - p\cdot |T_a| = \left(1-\frac{6p}{5}\right)\cdot (|T_b|-|T_a|) + \left(1-\frac{11p}{5}\right)\cdot |T_a|.
\end{equation*}
We choose $p = 5/11 < 2/3$ to eliminate the $|T_a|$ term on the RHS.
Using $|M^*_{YY}|=U_b$ and Lemma~\ref{lem:count2}, we can lower-bound $\E[\ALG\mid H_b]$ as
\begin{align*}
    \E[\ALG\mid H_b] &\ge |T_b| - \frac{5}{11}\cdot |T_a| + \frac{3}{11}\cdot (|M^*_{1Y}|+|M^*_{YY}|) + \frac{7}{55}\cdot U_b \\
    &\ge \frac{6}{11}\cdot \OPT + \frac{5}{11}\cdot (|T_b| - |T_a|) - \frac{3}{11}\cdot U_a + \frac{3}{11}\cdot |M^*_2| + \frac{7}{55}\cdot U_b \\
    &\ge \frac{6}{11}\cdot \OPT + \frac{1}{2}\cdot |M^*_2| - \frac{U_a}{22} - \frac{U_b}{10}.
\end{align*}

\subsection{\texorpdfstring{{A Lower Bound in Terms of Survival Probabilities}}{A Lower Bound in Terms of Survival Probabilities}}
\label{sec:global-survival}

Taking expectations over $T_b$ and the arrival times of its edges, we have
\begin{equation}
\label{eq:unconditional-global-bound}
    \E[\ALG] \ge \frac{6}{11}\cdot \OPT + \frac{1}{2}\cdot \E[|M^*_2|] - \frac{1}{22}\cdot \E[U_a] - \frac{1}{10}\cdot \E[U_b].
\end{equation}
We express the three expectations on the RHS using a single function.

For each edge $e=(u,v)\in M^*$, consider running \Greedy{} on $G-e$, with all remaining edges keeping their original arrival times.
For any $t\in [0,1]$, let $\Survival_e(t)$ be the probability that both $u,v$ are unmatched at time $t$ in this run, i.e., the \emph{survival probability} of $e$ up to time $t$.
Note that the event defining $\Survival_e(t)$ depends only on the arrival times of edges other than $e$, and is therefore independent of $t_e$.
Define
\begin{equation*}
    \Survival(t) = \frac{1}{\OPT}\cdot \sum_{e\in M^*} \Survival_e(t).
\end{equation*}

We first express $\E[U_t]$ using $\Survival(t)$.
Fix an edge $e=(u,v)\in M^*$ and a time $t\le b$.
The event that both $u,v$ are unmatched in $T_t$ is equivalent to the event that $t_e>t$ and both $u,v$ are unmatched at time $t$ by \Greedy{} on $G-e$.
When $t_e>t$, deleting $e$ does not change any decision before time $t$.
Conversely, if both $u,v$ are unmatched in $T_t$, then $e$ cannot have arrived earlier, since the shadow matching would have selected it.
By independence and the uniform distribution of $t_e$, the probability that both $u,v$ are unmatched in $T_t$ is $(1-t)\cdot \Survival_e(t)$.
Summing over $e\in M^*$, we have
\begin{equation}
\label{eq:expected-U}
    \E[U_t] = \OPT\cdot (1-t)\cdot \Survival(t).
\end{equation}

We next express $\E[|M^*_2|]$.
Conditional on $t_e=t\in [a,b)$, an edge $e=(u,v)\in M^*$ is selected by the shadow matching if and only if both $u$ and $v$ are unmatched right before $t$ by \Greedy{} on $G-e$.
Thus, we have
\begin{equation}
\label{eq:expected-M2}
    \Pr[e\in M^*_2] = \int_a^b \Survival_e(t)\,dt, \qquad \text{and} \qquad \E[|M^*_2|]=\OPT\cdot\int_a^b \Survival(t)\,dt.
\end{equation}

Substituting Equations~\eqref{eq:expected-U} and~\eqref{eq:expected-M2} into Inequality~\eqref{eq:unconditional-global-bound}, we establish the following lower bound on the competitive ratio:
\begin{equation}
\label{eq:ratio-lb}
     \frac{\E[\ALG]}{\OPT} \ge \frac{6}{11} +\frac{1}{2}\cdot \int_a^b \Survival(t)\,dt - \frac{1-a}{22}\cdot \Survival(a) - \frac{1-b}{10}\cdot \Survival(b).
\end{equation}

\section{Polynomial Expansion and the Competitive Ratio}\label{sec:poly}

The lower bound we derived in Inequality~\eqref{eq:ratio-lb} depends on $G$ through $\Survival(t)$.
In this section, we first show that $\Survival(t)$ admits a polynomial expansion in powers of $1-t$, with nonnegative coefficients summing to $1$.
This allows us to evaluate the bound one power at a time.
We then choose $a$ and $b$, and prove that the resulting bound is at least $0.5445$.

\subsection{\texorpdfstring{{Survival Probabilities as Convex Combinations of Powers}}{Survival Probabilities as Convex Combinations of Powers}}
\label{sec:survival-expansion}

We first consider the probability that an arbitrary set of
vertices remains unmatched.
The following lemma will apply to each function $\Survival_e(t)$
defined in Section~\ref{sec:global-survival}.

\begin{lemma}
\label{lem:poly-expansion}
Consider running \Greedy{} on a graph $J$ where edges have independent and uniform arrival times in $[0,1]$.
For any vertex set $A\subseteq V(J)$ and any time $t\in [0,1]$, let $P(J,A,t)$ be the probability that all vertices in $A$ are unmatched at $t$.
Then there exist coefficients $c_n\ge 0$, only finitely many of which are nonzero, so that
\begin{equation*}
    P(J,A,t) = \sum_{n\ge 1} c_n\cdot (1-t)^{n-1}, \qquad
    \sum_{n\ge 1} c_n = 1.
\end{equation*}
\end{lemma}
\begin{proof}
We prove the statement by induction on $|E(J)|$, for all choices of $A$.
If $E(J)=\varnothing$, then $P(J,A,t)=1$.
The statement holds trivially with $c_1=1$.

Suppose that $E(J)\ne \varnothing$.
If all vertices in $A$ are unmatched at time $t$, then either they remain unmatched until the end of the run (which happens w.p. $P(J,A,1)$), or there is an edge with an endpoint in $A$ that arrives at some time $s\in (t,1]$ and is matched by \Greedy{}.

For the second event, fix an edge $e$ with $V(e)\cap A\ne \varnothing$ and condition on its arrival time $t_e=s$.
Before time $s$, \Greedy{} makes the same decisions on $J$ and $J-e$.
Thus, $e$ is the first edge with an endpoint in $A$ matched by \Greedy{} if and only if all vertices in $A\cup V(e)$ are unmatched right before $s$ in \Greedy{} on $J-e$.
Integrating over the uniform arrival time of $e$ and summing over the possible choices of $e$, we obtain
\begin{equation}
\label{eq:survival-recursion}
    P(J,A,t) = P(J,A,1) + \sum_{e\in E(J): V(e)\cap A\ne \varnothing} \int_t^1 P(J-e,A\cup V(e),s)\,ds.
\end{equation}

Each graph $J-e$ has fewer edges than $J$.
By the induction hypothesis, each integrand of the RHS is a finite nonnegative combination of powers $(1-s)^{n-1}$.
Moreover,
\begin{equation*}
    \int_t^1 c\cdot (1-s)^{n-1}\,ds = \frac{c}{n}\cdot (1-t)^n.
\end{equation*}
Therefore, Equation~\eqref{eq:survival-recursion} expresses $P(J,A,t)$ as a finite nonnegative combination of powers of $1-t$, including the constant term $P(J,A,1)$.
Finally, at time $t=0$, we have
\begin{equation*}
    \sum_{n\ge 1} c_n = P(J,A,0)=1,
\end{equation*}
which completes the induction.
\end{proof}

For each $e=(u,v)\in M^*$, applying Lemma~\ref{lem:poly-expansion} with $J=G-e$ and $A=\{u,v\}$ gives the stated representation for $\Survival_e(t)$.
Averaging over $e\in M^*$ preserves both nonnegativity and the sum of the coefficients.
Hence, there exist coefficients $c_n\ge 0$, only finitely many of which are nonzero, so that
\begin{equation}
\label{eq:r-power-expansion}
    \Survival(t) = \sum_{n\ge 1} c_n\cdot (1-t)^{n-1},
    \qquad
    \sum_{n\ge 1} c_n = 1.
\end{equation}

\subsection{Optimizing the Phase Parameters}
\label{sec:analytic-bound}

We now substitute the representation of $\Survival(t)$ into Inequality~\eqref{eq:ratio-lb}.
For each integer $n\ge 1$, define
\begin{equation*}
    L_n(a,b) = \frac{6}{11} + \frac{(1-a)^n-(1-b)^n}{2n} - \frac{(1-a)^n}{22} - \frac{(1-b)^n}{10}.
\end{equation*}
Notice that
\begin{equation*}
    \int_a^b(1-t)^{n-1}\,dt = \frac{(1-a)^n-(1-b)^n}{n}.
\end{equation*}
Substituting Equation~\eqref{eq:r-power-expansion} into Inequality~\eqref{eq:ratio-lb} yields
\begin{equation}
\label{eq:mixture-bound}
    \frac{\E[\ALG]}{\OPT} \ge \sum_{n\ge 1} c_n\cdot L_n(a,b) \ge \inf_{n\ge 1}\{ L_n(a,b)\}.
\end{equation}

It remains to choose the parameters $a$ and $b$ to maximize $\inf_{n\ge 1}\{ L_n(a,b)\}$ over $0\le a<b\le 3/8$.
For every integer $n\ge 1$, we have
\begin{equation*}
    \frac{\partial L_n(a,b)}{\partial b} = \frac{n+5}{10}\cdot (1-b)^{n-1} > 0.
\end{equation*}
Thus, for any $n\ge 1$ and any fixed $a$, $L_n(a,b)$ is increasing in $b$.
Set $b=3/8$.
Observe that $L_1(a,3/8)$ is strictly decreasing in $a$, while
$L_{12}(a,3/8)$ is strictly increasing.
Let $a^*$ be the unique solution in $(0,3/8)$ to $L_1(a,3/8)=L_{12}(a,3/8)$.
We have $a^*=0.1769129524\ldots$.
In particular, $a^*>1/6$ and $L_1(a^*,3/8)<0.544586$.

We next verify that
$L_n(a^*,3/8)\ge L_1(a^*,3/8)$ for every integer $n\ge1$.
The cases $n=1$ and $n=12$ hold by the definition of $a^*$.
For the remaining $n$ between $2$ and $15$, we list the following lower bounds on $L_n(a^*,3/8)$.
Every bound exceeds $0.544586$, and hence exceeds $L_1(a^*,3/8)$.
\begin{center}
\renewcommand{\arraystretch}{1.15}
\setlength{\tabcolsep}{12pt}
\begin{tabular}{cc@{\qquad}cc}
\toprule
$n$ & Lower bound & $n$ & Lower bound \\
\midrule
2 & 0.547309 & 9  & 0.544942 \\
3 & 0.547940 & 10 & 0.544738 \\
4 & 0.547631 & 11 & 0.544627 \\
5 & 0.546986 & 13 & 0.544590 \\
6 & 0.546305 & 14 & 0.544628 \\
7 & 0.545715 & 15 & 0.544685 \\
8 & 0.545261 &    &          \\
\bottomrule
\end{tabular}
\end{center}

For $n\ge15$, write
\begin{equation*}
    L_n(a^*,3/8)=\frac{6}{11}-A_n-B_n,
\end{equation*}
where
\begin{equation*}
    A_n=\left(\frac{1}{22}-\frac{1}{2n}\right)(1-a^*)^n,
    \qquad
    B_n=\left(\frac{1}{10}+\frac{1}{2n}\right)
        \left(\frac58\right)^n.
\end{equation*}
The sequence $B_n$ is strictly decreasing.
Moreover, for $n\ge15$,
\begin{equation*}
    A_{n+1}<A_n
    \quad\Longleftrightarrow\quad
    a^*\cdot n\cdot (n-10)>11.
\end{equation*}
Since $a^*>1/6$ and $n\cdot (n-10)\ge 75$ for $n\ge 15$, we have
\begin{equation*}
    a^*\cdot n\cdot (n-10)>\frac{75}{6}>11.
\end{equation*}
Thus, $A_n$ is also strictly decreasing for $n\ge 15$,
so $L_n(a^*,3/8)$ is strictly increasing over this range.

Therefore, we have
\begin{equation*}
    \max_{0\le a<b\le3/8} \left\{\inf_{n\ge 1}\{ L_n(a,b)\}\right\} = L_1(a^*,3/8) =
    \frac{5}{8} - \frac{5}{11}\cdot a^* = 0.5445850216\ldots > 0.5445,
\end{equation*}
which concludes the proof of Theorem~\ref{thm:main}.

\section{\texorpdfstring{{Local Greedy Comparison}}{Local Greedy Comparison}}
\label{sec:local}

We first describe the remaining randomness after conditioning on $H_b$ and $B$ in Section~\ref{sec:cond-clocks}.
We then prove Lemma~\ref{lem:greedy-comparison} for the cases when $k_e=1$ and $k_e=2$ in Sections~\ref{sec:one-ref} and \ref{sec:two-ref}, respectively.

\subsection{Conditional Arrival Times}
\label{sec:cond-clocks}

Recall that $H_b$ records $T_b$ and the arrival times of its edges.
Conditional on $H_b$, we also fix an edge $e=(x_1,x_2)\in T_a$, and define $B$ to be the graph on $(X\cup Y)\setminus \{x_1,x_2\}$ containing all $YY$-edges and all $XY$-edges arriving after time $b$.
Let $\tau_v$ be the arrival time of the shadow edge matching $v$ for each $v\in V(T_b)$; let $\tau_v=+\infty$ for each $v\in Y$.

\begin{lemma}
\label{lem:cond-clocks}
Conditional on $H_b$, the arrival times of all edges outside $T_b$ are mutually independent.
For each edge $(u,v)\in E\setminus T_b$, we have
\begin{equation*}
    t_{(u,v)}\sim U\left(\min\{\tau_u,\tau_v,b\},1\right).
\end{equation*}
\end{lemma}
\begin{proof}
For fixed $H_b$, the arrival times of all remaining edges $(u,v)\in E\setminus T_b$ must satisfy
$t_{(u,v)}>\min\{\tau_u,\tau_v,b\}$.
These constraints also suffice to reproduce the same shadow matching and its arrival times.
Indeed, when processing edges in arrival order, every edge outside $T_b$ that arrives before $b$ has an endpoint already matched by an earlier shadow edge, so it is rejected; the recorded shadow edges are therefore accepted as before.
Thus, conditioning on $H_b$ imposes only these separate constraints on the remaining edges.
Their arrival times remain independent, and each is uniformly distributed on its allowed interval.
\end{proof}

Let the fixed edge $e=(x_1,x_2)$ arrive at time $s<a$.
We further condition on $B$.
Since the edges between $\{x_1,x_2\}$ and $Y$ are not used to determine $B$, Lemma~\ref{lem:cond-clocks} indicates that their arrival times remain mutually independent, each distributed as $U(s,1)$.
Hence, each such edge arrives after $b$ (or in other words, appears in $B_e$), independently w.p.
\begin{equation*}
    q = \frac{1-b}{1-s}\ge 1-b\ge \frac{5}{8}.
\end{equation*}
Conditional on which of these edges arrive after $b$, all edges of $B_e$ have independent arrival times distributed as $U(b,1)$.
We can thus generate the coupled matchings $M(B)$ and $M(B_e)$ as follows.

\paragraph{Equivalent formulation of the coupled matchings.}
Choose a uniform random order $\sigma$ of all edges of $B$ and all edges between $\{x_1,x_2\}$ and $Y$.
Independently, retain each edge between $\{x_1,x_2\}$ and $Y$ w.p. $q$.
To obtain $M(B_e)$, run \Greedy{} in the chosen order $\sigma$, skipping the edges between $\{x_1,x_2\}$ and $Y$ that are not retained.
To obtain $M(B)$, use the same order $\sigma$ but skip all edges between $\{x_1,x_2\}$ and $Y$.

\medskip

For the rest of this section, we keep $H_b$ and $B$ fixed and consider this equivalent formulation.
We write $k=k_e$ and omit the conditioning on $H_b$ and $B$ from the notation.
Our goal is to prove
\begin{equation}
\label{eq:comparision-bound}
    \E[3\N+2\R+\K]\ge 3k.
\end{equation}
Since $\N,\R,\K$ are nonnegative, the inequality is immediate when $k=0$.
It remains to consider the cases when $k=1$ and $k=2$.

\subsection{One Reference Vertex}
\label{sec:one-ref}

We first consider the graph with only one of $x_1,x_2$, say $x_1$, added to $B$.
Let $C=B_e-x_2$, and let $M(C)$ be its greedy matching under the edge order $\sigma$.
Let $\N_1\in \{0,1\}$ indicate whether $x_1$ is matched in $M(C)$, and let $\R_1$ count the reference vertices of $e$ matched in $M(C)$.
The following lemma allows us to add $x_2$ back while using the same bound on $C$.
We defer its proof to Appendix~\ref{Appendix:A}.

\begin{lemma}
\label{lem:vertex-insertion}
For every fixed edge order $\sigma$ and every choice of retained edges between $\{x_1,x_2\}$ and $Y$, we have
\begin{equation*}
    3\N+2\R\ge 3\N_1+2\R_1.
\end{equation*}
\end{lemma}

Now suppose that $k=1$, and let $y_1$ with $(x_1,y_1)\in M^*$ be the unique reference vertex of $e$.
In this case, $\R_1$ indicates whether $y_1$ is matched in $M(C)$, and $\K$ indicates whether it is matched in $M(B)$.
We first show that
\begin{equation}
\label{eq:one-reference-bound}
    \E[3\N_1+2\R_1+\K]\ge 3.
\end{equation}

Fix the edge order $\sigma$ and which edges incident to $x_1$, other than $(x_1,y_1)$, are retained.
The edge $(x_1,y_1)$ is retained w.p. $q$.
Let $C_0$ be the graph obtained by omitting $(x_1,y_1)$ from $C$.
Under this conditioning, $M(B)$ and $M(C_0)$ are fixed.
We distinguish three cases and consider the contribution to $3\N_1+2\R_1+\K$.
\begin{itemize}
    \item \textbf{$\K=1$.}
    If $\N_1=1$, the contribution is at least $3$.
    If $\N_1=0$, then no edge incident to $x_1$ is matched in $C$, and \Greedy{} makes the same decisions within $B$ as $M(B)$.
    We have $\R_1=1$, and $3\N_1+2\R_1+\K=3$.

    \item \textbf{$\K=0$ and $M(C_0)$ matches $x_1$.}
    Then $M(C)$ matches $x_1$ regardless of whether $(x_1,y_1)$ is retained or not.
    Either $(x_1,y_1)$ is matched in $M(C)$, or \Greedy{} makes the same decisions as on $C_0$.
    Thus $\N_1=1$, and the contribution is at least $3$.

    \item \textbf{$\K=0$ and $M(C_0)$ does not match $x_1$.}
    Since no edge incident to $x_1$ is accepted in $C_0$, its greedy matching within $B$ is $M(B)$.
    As $\K=0$, both $x_1$ and $y_1$ are unmatched in $M(C_0)$.
    If $(x_1,y_1)$ is retained, it gets matched in $M(C)$, giving $\N_1=\R_1=1$.
    If it is not retained, then $\N_1=\R_1=0$.
    The conditional expected contribution is $(2+3)\cdot q=5q \ge 25/8>3$.
\end{itemize}

In every case, the conditional expected contribution is
at least $3$.
Taking expectations over $\sigma$ and the retained edges other than $(x_1,y_1)$ proves Inequality~\eqref{eq:one-reference-bound}.
By Lemma~\ref{lem:vertex-insertion}, we have
\begin{equation*}
    \E[3\N+2\R+\K] \ge \E[3\N_1+2\R_1+\K] \ge 3,
\end{equation*}
which proves Lemma~\ref{lem:greedy-comparison} when $k=1$.

\subsection{Two Reference Vertices}
\label{sec:two-ref}

Let the reference vertices of $e$ be $y_1$ and $y_2$ so that $(x_1,y_1),(x_2,y_2)\in M^*$.
Applying Lemma~\ref{lem:greedy-comparison} for $k=1$ separately to $y_1$ and $y_2$ would count $3\N$ twice.
We instead condition on the first edge in $\sigma$.
For each possible first edge, we separate the contribution of the vertices it matches from the contribution still to be analyzed.
For $i\in \{1,2\}$, let $\mu_i$ be the probability that $y_i$ is matched in $M(B)$.
Thus, $\E[\K]=\mu_1+\mu_2$.

Fix the retention choices of all edges between $\{x_1,x_2\}$ and $Y\setminus\{y_1,y_2\}$.
Let $L$ count all edges of $B$, all retained edges from $\{x_1,x_2\}$ to $Y\setminus\{y_1,y_2\}$, and all edges from $\{x_1,x_2\}$ to $\{y_1,y_2\}$, regardless of whether retained or not.
We prove Lemma~\ref{lem:greedy-comparison} when $k=2$ by induction on $L\ge 2$.
When $L=2$, the only edges are $(x_1,y_1)$ and $(x_2,y_2)$.
Each contributes $5$ to the LHS of Inequality~\eqref{eq:comparision-bound} whenever it is retained.
We have
\begin{equation*}
    \E[3\N+2\R+\K]=10q\ge 6.
\end{equation*}

Suppose that $L>2$.
Assume that Lemma~\ref{lem:greedy-comparison} holds for $k=2$ whenever fewer than $L$ edges are counted, and both $(x_1,y_1),(x_2,y_2)$ exist before retention.
Consider the first of these $L$ edges in $\sigma$.
Every edge is equally likely to be first.
Conditional on the first edge, the order of the remaining edges is still uniform.
We first handle the following two cases.

\paragraph{The first edge $f$ belongs to $B$.}
Both $M(B_e)$ and $M(B)$ accept $f$.
Each reference vertex covered by $f$ contributes $2+1=3$ through $2\R+\K$.
If neither reference vertex is covered by $f$, the induction hypothesis gives an expected contribution of at least $6$.
If only one reference vertex is covered by $f$, Section~\ref{sec:one-ref} gives an expected contribution of at least $3$ from $x_1,x_2$ and the reference vertex not covered by $f$.
If both reference vertices are covered by $f$, the contribution is already $6$.
Thus, for every $f$ belonging to $B$, we have
\begin{equation*}
    \E[3\N+2\R+\K\mid f\text{ is the first}]\ge 6.
\end{equation*}

\paragraph{The first edge joins an endpoint of $e$ to a non-reference vertex.}
By symmetry, let this first edge be $(x_1,z)$, where $z\in Y\setminus\{y_1,y_2\}$.
It is among the $L$ edges and is retained.
Hence, $M(B_e)$ accepts it, and $x_1$ contributes $3$.
After matching $(x_1,z)$, \Greedy{} continues on $B_e-\{x_1,z\}$, which consists of $B-z$ together with $x_2$ and its retained edges to vertices other than $z$.
We therefore need the following lemma, whose proof is given in Appendix~\ref{Appendix:A}.

\begin{lemma}
\label{lem:one-ref-deletion}
For $i\in\{1,2\}$ and $z\in V(B)\setminus\{y_i\}$, let $D$ be the graph obtained from $B-z$ by adding $x_i$ and its retained edges to $Y\setminus\{z\}$.
Suppose that $(x_i,y_i)$ is one of the edges retained independently w.p. $q$.
Let $n$ and $r$ indicate whether $x_i$ and $y_i$ are matched by the greedy matching on $D$, respectively.
Then
\begin{equation*}
    \E[3n+2r]\ge 3-\mu_i.
\end{equation*}
\end{lemma}

By Lemma~\ref{lem:one-ref-deletion}, the expected contribution from $x_2,y_2$ is at least $3-\mu_2$.
Together with $\E[\K]=\mu_1+\mu_2$, we obtain
\begin{equation*}
    \E[3\N+2\R+\K\mid (x_1,z)\text{ is the first}] \ge 3+(3-\mu_2)+\mu_1+\mu_2 = 6+\mu_1\ge 6.
\end{equation*}

It remains to consider the case when an edge to a reference vertex is the first.
Besides $(x_1,y_1)$ and $(x_2,y_2)$, there may be neither, one, or both of $(x_1,y_2)$ and $(x_2,y_1)$.
We call $(x_1,y_2)$ and $(x_2,y_1)$ cross edges.

\paragraph{No cross edges.}
Assume w.l.o.g. that $(x_1,y_1)$ is the first.
If it is retained, $M(B_e)$ accepts it: $x_1$ and $y_1$ contribute $3+2=5$.
After accepting $(x_1,y_1)$, \Greedy{} continues on the graph consisting of $B-y_1$ together with $x_2$ and its retained edges.
By Lemma~\ref{lem:one-ref-deletion}, we have
\begin{equation*}
    \E[3\N+2\R+\K \mid (x_1,y_1)\text{ is the first and retained}] \ge 5+(3-\mu_2)+\mu_1+\mu_2 = 8+\mu_1.
\end{equation*}
If it is not retained, the following lemma gives a lower bound of $3+\mu_1$.

\begin{lemma}
\label{lem:one-ref-two}
If there exists an edge from $\{x_1,x_2\}$ to $y_i$ before retention, then
\begin{equation}
\label{eq:one-ref-two}
    \E[3\N+2\R+\K]\ge 3+\mu_{3-i}.
\end{equation}
\end{lemma}
\begin{proof}
Apply Inequality~\eqref{eq:one-reference-bound} with only $y_i$ counted as a reference vertex, and then use Lemma~\ref{lem:vertex-insertion} to include the other endpoint of $e$.
Vertices $x_1,x_2$ and $y_i$ contribute at least $3$ to the LHS of Inequality~\eqref{eq:one-ref-two}, while $y_{3-i}$ contributes at least $\mu_{3-i}$.
\end{proof}

Taking the expectation yields
\begin{equation*}
    \E[3\N+2\R+\K\mid (x_1,y_1)\text{ is the first}] \ge q\cdot (8+\mu_1)+(1-q)\cdot (3+\mu_1) = 3+5q+\mu_1\ge 6.
\end{equation*}

\paragraph{One cross edge.}
Assume w.l.o.g. that $(x_1,y_2)$ exists and
$(x_2,y_1)$ does not.
We list the conditional lower bounds on $\E[3\N+2\R+\K]$ for the three possible first edges.
\begin{center}
\renewcommand{\arraystretch}{1.15}
\setlength{\tabcolsep}{12pt}
\begin{tabular}{ccc}
    \toprule
    First edge & Retained & Not retained \\
    \midrule
    $(x_1,y_1)$
    & $8+\mu_1$
    & $4+\mu_1-\mu_2$ [Lemma~\ref{lem:shared-ref}] \\
    $(x_2,y_2)$
    & $8+\mu_2$
    & $3+\mu_2$ [Lemma~\ref{lem:one-ref-two}] \\
    $(x_1,y_2)$
    & $5+\mu_1+\mu_2$
    & $6$ [Induction hypothesis] \\
    \bottomrule
\end{tabular}
\end{center}

If $(x_1,y_1)$ or $(x_2,y_2)$ is the first and retained, the analysis from the previous case applies.
If $(x_1,y_2)$ is first and retained, accepting it contributes $5$; adding $\E[\K]=\mu_1+\mu_2$ gives the stated lower bound.

When $(x_1,y_1)$ is the first and not retained, we need an additional lower bound.
For this case, among the remaining $L-1$ edges, neither $x_1$ nor $x_2$ has an edge connecting to $y_1$.

\begin{lemma}
\label{lem:shared-ref}
Assume that the only edges between $\{x_1,x_2\}$ and $\{y_1,y_2\}$ before retention are $(x_1,y_2)$ and $(x_2,y_2)$.
Then we have
\begin{equation*}
    \E[3\N+2\R+\K]\ge 4+\mu_1-\mu_2.
\end{equation*}
\end{lemma}
We defer the proof to Appendix~\ref{Appendix:A}.

If $(x_2,y_2)$ is not retained, then Lemma~\ref{lem:one-ref-two} applies.
If $(x_1,y_2)$ is not retained, the remaining $L-1$ edges still contain $(x_1,y_1),(x_2,y_2)$, so the induction hypothesis applies.

Each of the other $L-3$ possible first edges has conditional expected contribution at least $6$, as shown before.
Consequently,
\begin{align*}
    \E[3\N+2\R+\K]
    &\ge \frac{1}{L}\left(6\cdot (L-3) + q\cdot (21+2\mu_1+2\mu_2) + (1-q)\cdot (13+\mu_1)\right) \\
    &\ge \frac{6\cdot (L-3)+13+8q}{L} \ge 6,
\end{align*}
where the last inequality uses $q\ge 5/8$.

\paragraph{Two cross edges.}
All four edges between $\{x_1,x_2\}$ and $\{y_1,y_2\}$ exist.
Assume w.l.o.g. that $(x_1,y_1)$ is the first.
If it is retained, the same calculation as in the case with no cross edges gives a lower bound of $8+\mu_1\ge6$.
If it is not retained, the remaining $L-1$ edges contain $(x_1,y_2),(x_2,y_1)$.
Interchange the labels $y_1$ and $y_2$.
The induction hypothesis then gives a lower bound of $6$.

\medskip

We have finished the induction.
Taking the expectation over the retained edges between $\{x_1,x_2\}$ and $Y\setminus\{y_1,y_2\}$ proves Lemma~\ref{lem:greedy-comparison} when $k=2$.

\clearpage
\section{Upper Bounds on Graphs Known up to Isomorphism}
\label{sec:upper-proof}

The thick-$\mathcal{Z}$ construction of \citet{conf/ipco/GuruganeshS17} and the triangle with three leaves of \citet{FavrholdtVatshelleManuscript} use the same approach: choose a fixed graph and bound what the algorithm can gain from each possible decision. The four-vertex graph below illustrates this argument.

\begin{figure}[H]
\centering
\begin{minipage}{.3\linewidth}
\centering
\begin{tikzpicture}[scale=.9,
 vertex/.style={circle,draw=black,fill=black,inner sep=1.45pt,line width=.45pt},
 ordinary/.style={black,line width=.5pt},
 matching/.style={figureblue,line width=1.2pt}]
\coordinate (a) at (0,1.2);
\coordinate (b) at (0,.35);
\coordinate (c) at (-1,-.95);
\coordinate (d) at (1,-.95);
\draw[ordinary] (c)--(b)--(d);
\draw[matching] (a)--(b) (c)--(d);
\foreach \v in {a,b,c,d} \node[vertex] at (\v) {};
\end{tikzpicture}
\caption{The four-vertex graph.}\label{fig:upper-warmup}
\end{minipage}\hfill
\begin{minipage}{.66\linewidth}
\newcommand{\warmupblue}[1]{\tikz[baseline=(X.base)]{\node[inner sep=0pt,text=figureblue](X){#1};\draw[figureblue,line width=1.2pt]([yshift=-2.3pt]X.base west)--([yshift=-2.3pt]X.base east);}}
The two \warmupblue{blue edges} form the unique maximum matching. Random labels hide whether the first edge is blue, which happens w.p. $1/2$. Accepting a black edge or rejecting a blue edge limits the final matching to one edge; in the other case it has at most two. Thus either first decision gives expected size at most $(1+2)/2=3/2$. \Greedy{} attains this value, so the optimal ratio on this graph is $(3/2)/2=3/4$.
\end{minipage}
\end{figure}
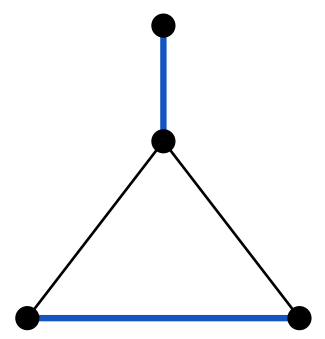

We formalize this reasoning as an online decision problem on a fixed graph $G$ with $n$ vertices, $m$ edges, and maximum matching size $\OPT(G)>0$. The algorithm is given $G$ up to isomorphism, a uniformly random bijection $\pi\colon V(G)\to[n]=\{0,\dots,n-1\}$ assigns vertex labels, and the edges of $\pi(G)$ arrive in an independent uniformly random order. Write $V^\star(G)$ for the maximum expected matching size achievable by an online algorithm in this problem. Averaging over $\pi$ shows that every online algorithm has some fixed labeling of $G$ on which its expected matching size is at most $V^\star(G)$, so no online algorithm achieves a competitive ratio greater than $V^\star(G)/\OPT(G)$.

\subsection{\texorpdfstring{{Dynamic Program and Optimal Policy}}{Dynamic Program and Optimal Policy}}
\label{sec:dynamic-program}
At any step, let $H$ be the subgraph of edges observed so far and $U\subseteq[n]$ be the set of currently unmatched labels. Let
\[
 C(H)=\bigl|\{\pi\colon V(G)\to[n]\text{ bijective}:E(H)\subseteq\pi(E(G))\}\bigr|
\]
be the number of labelings compatible with $H$. When $k=|E(H)|<m$, the next arriving edge $e\notin E(H)$ has distribution
\begin{equation}\label{eq:fixed-graph-transition}
 p(e\mid H)=\frac{C(H+e)}{(m-k)C(H)}.
\end{equation}
Define $V(H,U)$ backwards from $V(H,U)=0$ for $|E(H)|=m$ by
\begin{equation}\label{eq:fixed-graph-bellman}
V(H,U)=\sum_{\substack{e\notin E(H)\\C(H+e)>0}}p(e\mid H)
\begin{cases}
\max\{V(H+e,U),\;1+V(H+e,U\setminus e)\},&e\subseteq U,\\
V(H+e,U),&e\not\subseteq U,
\end{cases}
\end{equation}
where $U\setminus e$ removes both endpoints of $e$.

\begin{lemma}\label{lem:fixed-graph-optimal}
For every graph $G$, $V^\star(G)=V(\varnothing,[n])$, attained by accepting each feasible arriving edge $e\subseteq U$ if and only if $1+V(H+e,U\setminus e)\ge V(H+e,U)$.
\end{lemma}
\begin{proof}
Any ordered prefix of $k$ edges has probability $1/(m(m-1)\cdots(m-k+1))$ under every compatible labeling and zero under all others. Independent of past algorithm choices and of the arrival order within $E(H)$, the posterior over $\pi$ is therefore uniform on the $C(H)$ compatible labelings. Each of the $C(H+e)$ compatible labelings containing $e$ presents $e$ next w.p. $1/(m-k)$, giving Equation~\eqref{eq:fixed-graph-transition}. Since $H$ determines the distribution of future arrivals and $U$ determines feasibility, $(H,U)$ is a sufficient state, and backward induction on $m-|E(H)|$ yields Equation~\eqref{eq:fixed-graph-bellman}.
\end{proof}

\subsection{\texorpdfstring{{Upper Bound for General Graphs}}{Upper Bound for General Graphs}}
\label{sec:general-upper-bound}
\begin{figure}[t]
  \centering
\begin{tikzpicture}[
  vertex/.style={circle,draw=black,fill=black,inner sep=1.45pt,line width=.45pt},
  ordinary/.style={black,line width=.5pt},
  matching/.style={figureblue,line width=1.2pt},
  missing/.style={red!70!black,dashed,line width=.8pt},
  every label/.style={font=\footnotesize,inner sep=1.5pt},
  title/.style={font=\large}
]
\begin{scope}
  \node[title,align=center,anchor=south] at (0,1.75) {$G_4$\\{\small $0.75$}};
  \coordinate (0) at (0,1.2);
  \coordinate (3) at (0,.35);
  \coordinate (1) at (-1,-.95);
  \coordinate (2) at (1,-.95);
  \draw[ordinary] (1)--(3)--(2);
  \draw[matching] (0)--(3) (1)--(2);
  \node[vertex] at (0) {};
  \node[vertex] at (3) {};
  \node[vertex] at (1) {};
  \node[vertex] at (2) {};
\end{scope}
\begin{scope}[xshift=7.2cm]
  \node[title,align=center,anchor=south] at (0,1.75) {$G_6$\\{\small $0.722369$}};
  \coordinate (1) at (-.9,.55);
  \coordinate (2) at (.9,.55);
  \coordinate (5) at (-1.5,1.3);
  \coordinate (4) at (1.5,1.3);
  \coordinate (0) at (-.7,-1);
  \coordinate (3) at (.7,-1);
  \draw[ordinary] (1)--(2)--(5);
  \foreach \u in {0,3}
    \foreach \v in {1,2} \draw[ordinary] (\u)--(\v);
  \draw[missing] (1)--(4);
  \draw[matching] (0)--(3) (1)--(5) (2)--(4);
  \foreach \v in {4,5} \node[vertex] at (\v) {};
  \foreach \v in {0,3} \node[vertex] at (\v) {};
  \node[vertex] at (1) {};
  \node[vertex] at (2) {};
\end{scope}
\begin{scope}[yshift=-4.8cm]
  \node[title,align=center,anchor=south] at (0,1.75) {$G_8$\\{\small $0.708856$}};
  \coordinate (6) at (-1.05,.78);
  \coordinate (7) at (1.05,.78);
  \coordinate (2) at (-1.95,1.22);
  \coordinate (3) at (1.95,1.22);
  \coordinate (0) at (-2.3,-.08);
  \coordinate (4) at (-2.3,-1);
  \coordinate (1) at (2.3,-.08);
  \coordinate (5) at (2.3,-1);
  \draw[ordinary] (6)--(7);
  \foreach \u in {0,4,1,5}
    \foreach \v in {6,7} \draw[ordinary] (\u)--(\v);
  \draw[matching] (0)--(4) (1)--(5) (2)--(6) (3)--(7);
  \foreach \v in {6,7} \node[vertex] at (\v) {};
  \foreach \v in {0,2,4} \node[vertex] at (\v) {};
  \foreach \v in {1,3,5} \node[vertex] at (\v) {};
\end{scope}
\begin{scope}[xshift=7.2cm,yshift=-4.8cm]
  \node[title,align=center,anchor=south] at (0,1.75) {$G_{10}$\\{\small $0.696327$}};
  \coordinate (7) at (-1.05,.78);
  \coordinate (8) at (1.05,.78);
  \coordinate (9) at (0,-.52);
  \coordinate (2) at (-1.95,1.22);
  \coordinate (3) at (1.95,1.22);
  \coordinate (4) at (0,-1.28);
  \coordinate (6) at (-2.3,-.08);
  \coordinate (1) at (-2.3,-1);
  \coordinate (0) at (2.3,-.08);
  \coordinate (5) at (2.3,-1);
  \draw[ordinary] (7)--(8)--(9)--(7);
  \foreach \v in {8,9} \draw[ordinary] (6)--(\v);
  \foreach \u in {1,0,5}
    \foreach \v in {7,8,9} \draw[ordinary] (\u)--(\v);
  \draw[missing] (6)--(7);
  \draw[matching] (0)--(5) (1)--(6) (2)--(7) (3)--(8) (4)--(9);
  \foreach \v in {7,8} \node[vertex] at (\v) {};
  \foreach \v in {1,2,6} \node[vertex] at (\v) {};
  \foreach \v in {0,3,5} \node[vertex] at (\v) {};
  \node[vertex] at (4) {};
  \node[vertex] at (9) {};
\end{scope}
\end{tikzpicture}
\centering
  \caption{Worst-case graphs $G_n$ on at most $n$ vertices. \\ \protect\tikz[baseline=(X.base)]{\protect\node[inner sep=0pt,text=figureblue](X){Blue edges};\protect\draw[figureblue,line width=1.2pt]([yshift=-2.3pt]X.base west)--([yshift=-2.3pt]X.base east);} form the (unique) maximum matchings. \\ \protect\tikz[baseline=(Y.base)]{\protect\node[inner sep=0pt,text=red!70!black](Y){Red dashed edges};\protect\draw[red!70!black,dashed,line width=.8pt]([yshift=-2.3pt]Y.base west)--([yshift=-2.3pt]Y.base east);} in $G_6$ and $G_{10}$ are absent, breaking their symmetries.}\label{fig:upper-family}
\end{figure}
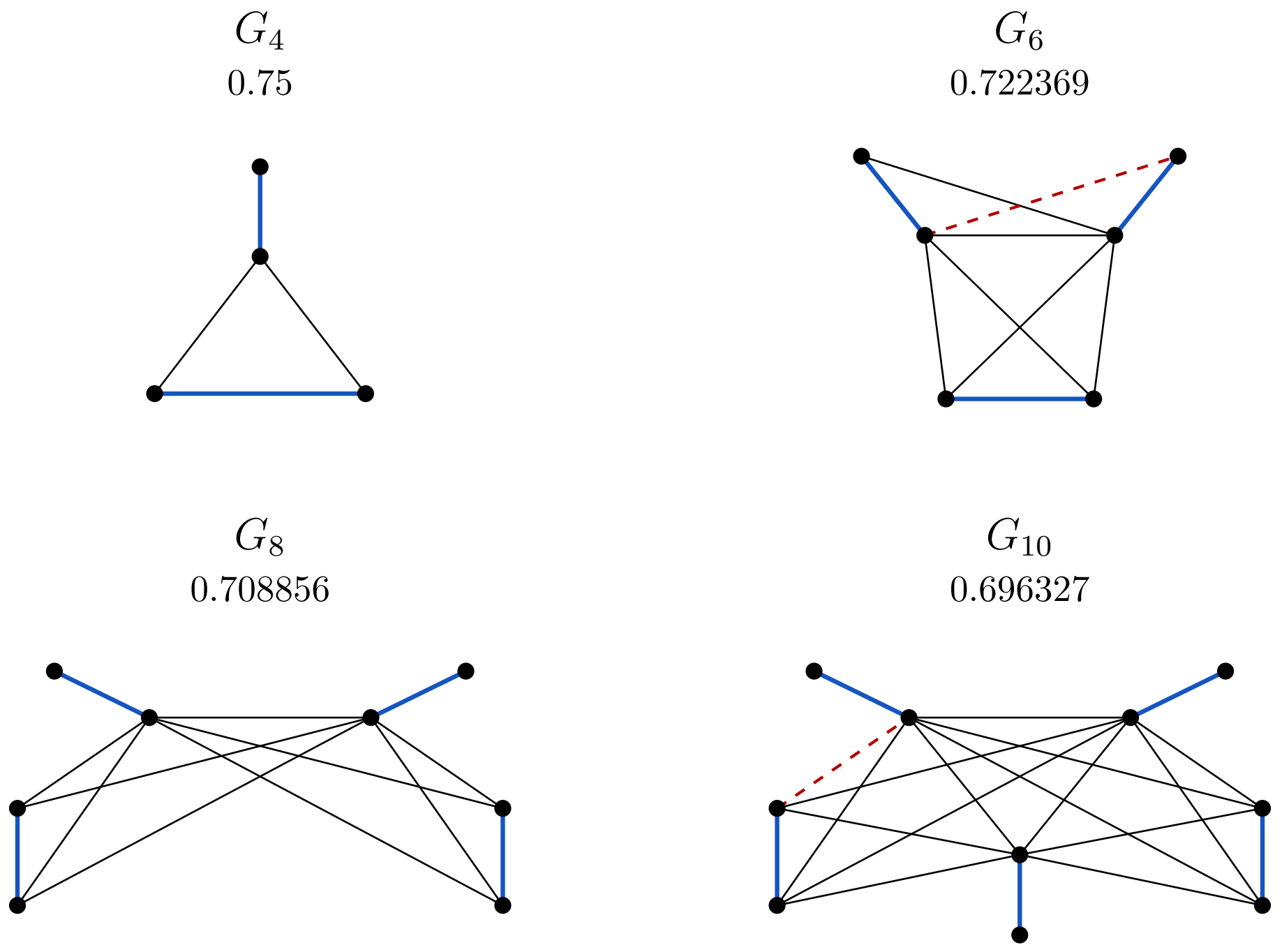

Evaluating Equation~\eqref{eq:fixed-graph-bellman} on the graphs in Figure~\ref{fig:upper-family} gives the ratios displayed beneath their names.

\begin{proof}[Proof of Theorem~\ref{thm:upper-bound}]
The $10$-vertex, $19$-edge graph $G_{10}$ in Figure~\ref{fig:upper-family} has $\OPT(G_{10})=5$. Exact rational evaluation of Equation~\eqref{eq:fixed-graph-bellman} gives
\[
  \frac{V^\star(G_{10})}{\OPT(G_{10})}=\frac{2941136911678289}{4223788208640000}<0.6964.\qedhere
\]
\end{proof}

\subsection{\texorpdfstring{{Upper Bound for Bipartite Graphs}}{Upper Bound for Bipartite Graphs}}
\label{sec:bipartite-upper-bound}
\begin{figure}[t]
\centering
\begin{tikzpicture}[x=1cm,y=1cm,
 vertex/.style={circle,fill=black,inner sep=1.3pt},
 cross/.style={black,line width=.4pt},
 matching/.style={figureblue,line width=1pt},
 every node/.style={font=\footnotesize}]
\node[font=\small\bfseries] at (4.4,1.4) {Optimal for $n$};
\node[font=\small\bfseries] at (11.3,1.4) {Best found};
\draw[black,line width=.4pt] (9.25,1.05)--(9.25,-5.95);
\begin{scope}[shift={(0,0)}]
\node[align=center,anchor=south] at (1.2,.32) {$n=4: B_{1,1}$\\$0.833333$};
\fill[figureblue!6] (-.30,-0.5400) rectangle (2.70,-0.8600);
\fill[green!6] (-.30,-1.2400) rectangle (2.70,-1.5600);
\draw[cross] (0,-0.7000)--(2.4,-1.4000);
\draw[matching] (0,-0.7000)--(2.4,-0.7000);
\node[vertex] at (0,-0.7000) {}; \node[vertex] at (2.4,-0.7000) {};
\draw[matching] (0,-1.4000)--(2.4,-1.4000);
\node[vertex] at (0,-1.4000) {}; \node[vertex] at (2.4,-1.4000) {};
\end{scope}
\begin{scope}[shift={(3.2,0)}]
\node[align=center,anchor=south] at (1.2,.32) {$n=6: B_{1,1,1}$\\$0.777778$};
\fill[figureblue!6] (-.30,-0.1900) rectangle (2.70,-0.5100);
\fill[green!6] (-.30,-0.8900) rectangle (2.70,-1.2100);
\fill[orange!10] (-.30,-1.5900) rectangle (2.70,-1.9100);
\draw[cross] (0,-0.3500)--(2.4,-1.0500);
\draw[cross] (0,-0.3500)--(2.4,-1.7500);
\draw[cross] (0,-1.0500)--(2.4,-1.7500);
\draw[matching] (0,-0.3500)--(2.4,-0.3500);
\node[vertex] at (0,-0.3500) {}; \node[vertex] at (2.4,-0.3500) {};
\draw[matching] (0,-1.0500)--(2.4,-1.0500);
\node[vertex] at (0,-1.0500) {}; \node[vertex] at (2.4,-1.0500) {};
\draw[matching] (0,-1.7500)--(2.4,-1.7500);
\node[vertex] at (0,-1.7500) {}; \node[vertex] at (2.4,-1.7500) {};
\end{scope}
\begin{scope}[shift={(6.4,0)}]
\node[align=center,anchor=south] at (1.2,.32) {$n=8: B_{1,1,1,1}$\\$0.757524$};
\fill[figureblue!6] (-.30,0.1600) rectangle (2.70,-0.1600);
\fill[green!6] (-.30,-0.5400) rectangle (2.70,-0.8600);
\fill[orange!10] (-.30,-1.2400) rectangle (2.70,-1.5600);
\fill[violet!7] (-.30,-1.9400) rectangle (2.70,-2.2600);
\draw[cross] (0,0.0000)--(2.4,-0.7000);
\draw[cross] (0,0.0000)--(2.4,-1.4000);
\draw[cross] (0,0.0000)--(2.4,-2.1000);
\draw[cross] (0,-0.7000)--(2.4,-1.4000);
\draw[cross] (0,-0.7000)--(2.4,-2.1000);
\draw[cross] (0,-1.4000)--(2.4,-2.1000);
\draw[matching] (0,0.0000)--(2.4,0.0000);
\node[vertex] at (0,0.0000) {}; \node[vertex] at (2.4,0.0000) {};
\draw[matching] (0,-0.7000)--(2.4,-0.7000);
\node[vertex] at (0,-0.7000) {}; \node[vertex] at (2.4,-0.7000) {};
\draw[matching] (0,-1.4000)--(2.4,-1.4000);
\node[vertex] at (0,-1.4000) {}; \node[vertex] at (2.4,-1.4000) {};
\draw[matching] (0,-2.1000)--(2.4,-2.1000);
\node[vertex] at (0,-2.1000) {}; \node[vertex] at (2.4,-2.1000) {};
\end{scope}
\begin{scope}[shift={(1.6,-3.5)}]
\node[align=center,anchor=south] at (1.2,.32) {$n=10: B_{2,1,2}$\\$0.739068$};
\fill[figureblue!6] (-.30,-0.0820) rectangle (2.70,-0.7580);
\fill[green!6] (-.30,-0.9220) rectangle (2.70,-1.1780);
\fill[orange!10] (-.30,-1.3420) rectangle (2.70,-2.0180);
\draw[cross] (0,-0.2100)--(2.4,-1.0500);
\draw[cross] (0,-0.2100)--(2.4,-1.4700);
\draw[cross] (0,-0.2100)--(2.4,-1.8900);
\draw[cross] (0,-0.6300)--(2.4,-1.0500);
\draw[cross] (0,-0.6300)--(2.4,-1.4700);
\draw[cross] (0,-0.6300)--(2.4,-1.8900);
\draw[cross] (0,-1.0500)--(2.4,-1.4700);
\draw[cross] (0,-1.0500)--(2.4,-1.8900);
\draw[matching] (0,-0.2100)--(2.4,-0.2100);
\node[vertex] at (0,-0.2100) {}; \node[vertex] at (2.4,-0.2100) {};
\draw[matching] (0,-0.6300)--(2.4,-0.6300);
\node[vertex] at (0,-0.6300) {}; \node[vertex] at (2.4,-0.6300) {};
\draw[matching] (0,-1.0500)--(2.4,-1.0500);
\node[vertex] at (0,-1.0500) {}; \node[vertex] at (2.4,-1.0500) {};
\draw[matching] (0,-1.4700)--(2.4,-1.4700);
\node[vertex] at (0,-1.4700) {}; \node[vertex] at (2.4,-1.4700) {};
\draw[matching] (0,-1.8900)--(2.4,-1.8900);
\node[vertex] at (0,-1.8900) {}; \node[vertex] at (2.4,-1.8900) {};
\end{scope}
\begin{scope}[shift={(4.8,-3.5)}]
\node[align=center,anchor=south] at (1.2,.32) {$n=12: B_{2,2,2}$\\$0.721443$};
\fill[figureblue!6] (-.30,0.1470) rectangle (2.70,-0.5670);
\fill[green!6] (-.30,-0.6930) rectangle (2.70,-1.4070);
\fill[orange!10] (-.30,-1.5330) rectangle (2.70,-2.2470);
\draw[cross] (0,0.0000)--(2.4,-0.8400);
\draw[cross] (0,0.0000)--(2.4,-1.2600);
\draw[cross] (0,0.0000)--(2.4,-1.6800);
\draw[cross] (0,0.0000)--(2.4,-2.1000);
\draw[cross] (0,-0.4200)--(2.4,-0.8400);
\draw[cross] (0,-0.4200)--(2.4,-1.2600);
\draw[cross] (0,-0.4200)--(2.4,-1.6800);
\draw[cross] (0,-0.4200)--(2.4,-2.1000);
\draw[cross] (0,-0.8400)--(2.4,-1.6800);
\draw[cross] (0,-0.8400)--(2.4,-2.1000);
\draw[cross] (0,-1.2600)--(2.4,-1.6800);
\draw[cross] (0,-1.2600)--(2.4,-2.1000);
\draw[matching] (0,0.0000)--(2.4,0.0000);
\node[vertex] at (0,0.0000) {}; \node[vertex] at (2.4,0.0000) {};
\draw[matching] (0,-0.4200)--(2.4,-0.4200);
\node[vertex] at (0,-0.4200) {}; \node[vertex] at (2.4,-0.4200) {};
\draw[matching] (0,-0.8400)--(2.4,-0.8400);
\node[vertex] at (0,-0.8400) {}; \node[vertex] at (2.4,-0.8400) {};
\draw[matching] (0,-1.2600)--(2.4,-1.2600);
\node[vertex] at (0,-1.2600) {}; \node[vertex] at (2.4,-1.2600) {};
\draw[matching] (0,-1.6800)--(2.4,-1.6800);
\node[vertex] at (0,-1.6800) {}; \node[vertex] at (2.4,-1.6800) {};
\draw[matching] (0,-2.1000)--(2.4,-2.1000);
\node[vertex] at (0,-2.1000) {}; \node[vertex] at (2.4,-2.1000) {};
\end{scope}
\begin{scope}[shift={(10.1,-0.9)}]
\node[align=center,anchor=south] at (1.2,.32) {$n=16: B_{2,3,3}$\\$0.704284$};
\fill[figureblue!6] (-.30,0.1600) rectangle (2.70,-0.6886);
\fill[green!6] (-.30,-0.8971) rectangle (2.70,-2.2743);
\fill[orange!10] (-.30,-2.4829) rectangle (2.70,-3.8600);
\draw[cross] (0,0.0000)--(2.4,-1.0571);
\draw[cross] (0,0.0000)--(2.4,-1.5857);
\draw[cross] (0,0.0000)--(2.4,-2.1143);
\draw[cross] (0,0.0000)--(2.4,-2.6429);
\draw[cross] (0,0.0000)--(2.4,-3.1714);
\draw[cross] (0,0.0000)--(2.4,-3.7000);
\draw[cross] (0,-0.5286)--(2.4,-1.0571);
\draw[cross] (0,-0.5286)--(2.4,-1.5857);
\draw[cross] (0,-0.5286)--(2.4,-2.1143);
\draw[cross] (0,-0.5286)--(2.4,-2.6429);
\draw[cross] (0,-0.5286)--(2.4,-3.1714);
\draw[cross] (0,-0.5286)--(2.4,-3.7000);
\draw[cross] (0,-1.0571)--(2.4,-2.6429);
\draw[cross] (0,-1.0571)--(2.4,-3.1714);
\draw[cross] (0,-1.0571)--(2.4,-3.7000);
\draw[cross] (0,-1.5857)--(2.4,-2.6429);
\draw[cross] (0,-1.5857)--(2.4,-3.1714);
\draw[cross] (0,-1.5857)--(2.4,-3.7000);
\draw[cross] (0,-2.1143)--(2.4,-2.6429);
\draw[cross] (0,-2.1143)--(2.4,-3.1714);
\draw[cross] (0,-2.1143)--(2.4,-3.7000);
\draw[matching] (0,0.0000)--(2.4,0.0000);
\node[vertex] at (0,0.0000) {}; \node[vertex] at (2.4,0.0000) {};
\draw[matching] (0,-0.5286)--(2.4,-0.5286);
\node[vertex] at (0,-0.5286) {}; \node[vertex] at (2.4,-0.5286) {};
\draw[matching] (0,-1.0571)--(2.4,-1.0571);
\node[vertex] at (0,-1.0571) {}; \node[vertex] at (2.4,-1.0571) {};
\draw[matching] (0,-1.5857)--(2.4,-1.5857);
\node[vertex] at (0,-1.5857) {}; \node[vertex] at (2.4,-1.5857) {};
\draw[matching] (0,-2.1143)--(2.4,-2.1143);
\node[vertex] at (0,-2.1143) {}; \node[vertex] at (2.4,-2.1143) {};
\draw[matching] (0,-2.6429)--(2.4,-2.6429);
\node[vertex] at (0,-2.6429) {}; \node[vertex] at (2.4,-2.6429) {};
\draw[matching] (0,-3.1714)--(2.4,-3.1714);
\node[vertex] at (0,-3.1714) {}; \node[vertex] at (2.4,-3.1714) {};
\draw[matching] (0,-3.7000)--(2.4,-3.7000);
\node[vertex] at (0,-3.7000) {}; \node[vertex] at (2.4,-3.7000) {};
\end{scope}
\end{tikzpicture}
\caption{Bipartite constructions and their ratios $V^\star(B)/\OPT(B)$. \\ Left: worst-case bipartite graphs identified by exhaustive search. \\ Right: our best-found sixteen-vertex construction.}
\label{fig:bipartite-construction}
\end{figure}
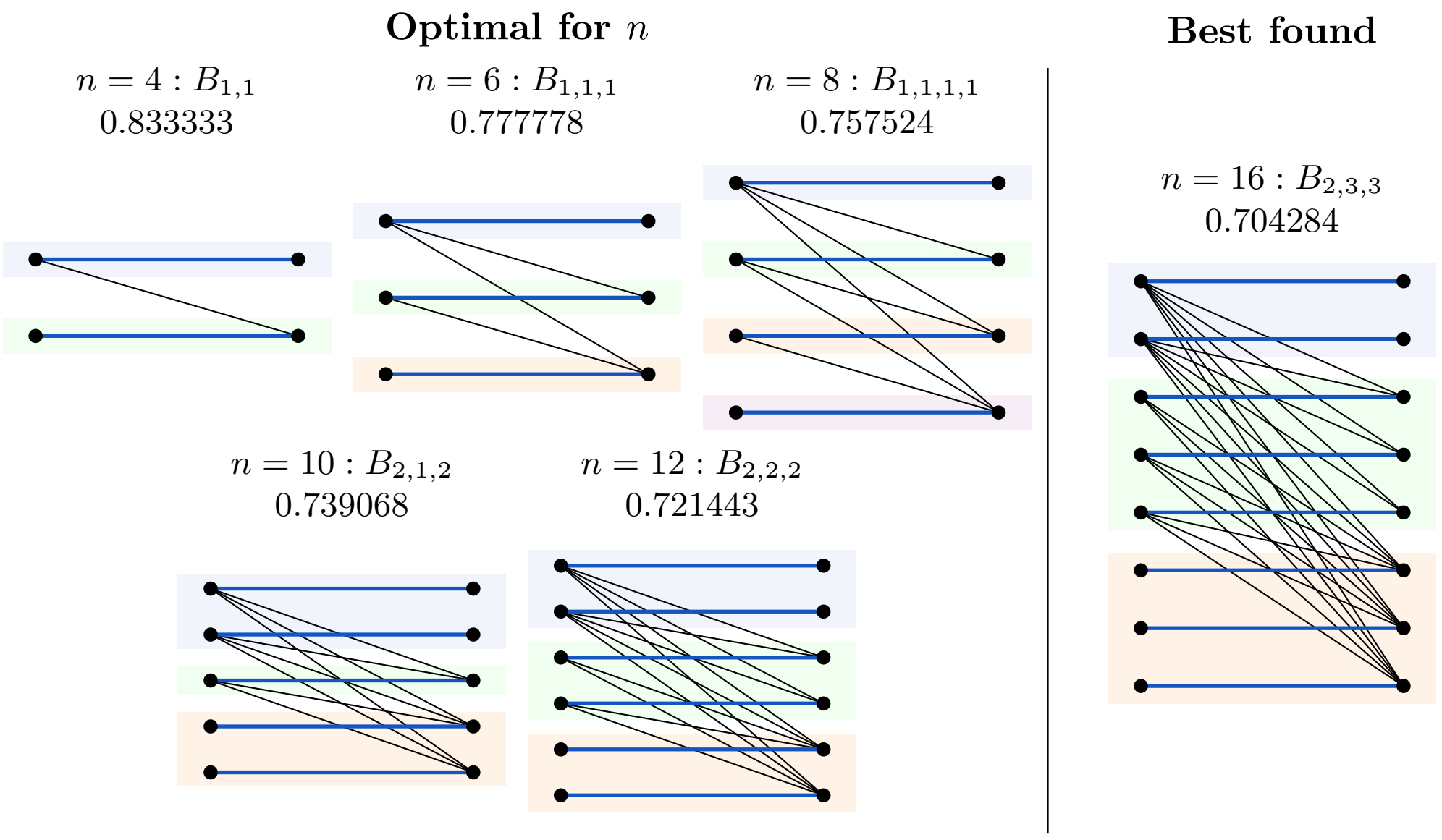

Unlike the worst-case general graphs, where $G_6$ and $G_{10}$ omit a single edge to break symmetry and follow no obvious parametric family, exhaustive search over all nonisomorphic bipartite graphs on $n=4,6,8,10,12$ vertices reveals a regular pattern (Figure~\ref{fig:bipartite-construction}, left). Define $B_{a_1,\ldots,a_r}$ by grouping disjoint matching edges into ordered blocks of sizes $a_1,\ldots,a_r$, then adding every edge from a left endpoint in an earlier block to a right endpoint in a later block. The original edges form the unique perfect matching, as matching the blocks in order shows. Searching larger members of this family yields the sixteen-vertex graph $B_{2,3,3}$ (Figure~\ref{fig:bipartite-construction}, right).

\begin{proof}[Proof of Theorem~\ref{thm:bipartite-upper}]
The graph $B_{2,3,3}$ in Figure~\ref{fig:bipartite-construction} has $16$ vertices, $29$ edges, and $\OPT(B_{2,3,3})=8$. Exact rational evaluation of Equation~\eqref{eq:fixed-graph-bellman} gives
\[
 \frac{V^\star(B_{2,3,3})}{\OPT(B_{2,3,3})}
 =\frac{3970969041694879055111207}{5638303485447723417600000}
 <0.7043.\qedhere
\]
\end{proof}

\paragraph{Extension to larger $n$.}
One might ask whether simply scaling up the block sizes, such as considering $B_{k,k,k}$ for large $k$, yields a strong asymptotic upper bound.
Unfortunately, this is not the case: like the sunflower graph in Figure~\ref{fig:sunflower}, $B_{k,k,k}$ has $\Theta(n^2)$ cross-block edges and only $n/2$ matching edges, so an online algorithm can achieve a competitive ratio of $1-O(1/\log n)$ by rejecting early edges.
Scaling to larger $n$ therefore requires a more careful construction.

\subsection{Implementation and Verification}
\label{sec:implementation}
We evaluate Equation~\eqref{eq:fixed-graph-bellman} with three implementations\footnote{Our code is available at \url{https://github.com/fstqwq/Random-Order-Matching-Upper-Bounds}.}, each serving a different purpose; Table~\ref{tab:implementations} compares their running times.

\paragraph{Python reference.}
A concise Python program evaluates Equation~\eqref{eq:fixed-graph-bellman} in exact rational arithmetic, memoizing states $(H,U)$ and using \texttt{nauty}~\cite{journals/jsc/McKayP14} to identify states up to consistent vertex relabeling.
Although its worst-case running time is $O(2^{m+n}\,n!\,\mathrm{poly}(n))$, far fewer states are reached in practice:
on $G_{10}$, the Python program visits $10\,447$ isomorphism classes of observed subgraphs and finishes in about two minutes.

\paragraph{Lean verification.}
To certify this computation, we wrote a second solver in Lean~4 and proved a theorem about its output. Fix a nonempty graph $G$, and average over all $n!$ labelings $\pi$ and all $m!$ arrival orders of $\pi(G)$. The theorem states that for every online strategy, deterministic or randomized with rational probabilities, the expected number of accepted edges divided by $\OPT(G)$ is at most the value returned by the solver. The proof first establishes this bound for the unreduced recurrence~\eqref{eq:fixed-graph-bellman}, and then shows that relabeling states by any vertex permutation, as the solver does, leaves the value unchanged. The Lean solver is much slower than the Python program; on $G_{10}$, it reproduces the exact value in Theorem~\ref{thm:upper-bound}.

\paragraph{C++ solver.}
For larger graphs, a multithreaded C++ solver evaluates Equation~\eqref{eq:fixed-graph-bellman} in exact arithmetic by parallel memoized search using a shared memoization table. In Lean~4, we prove the correctness of the underlying recurrence and memoization invariants: whenever each subproblem call or cache hit returns its true value, each step computes the exact value regardless of the vertex relabeling or summation order. The C++ solver implements this verified recurrence directly\footnote{Under standard runtime and external library assumptions documented in the repository, e.g., the correctness of the concurrent hash map.}; all three solvers also return identical exact ratios on $G_4, G_6, G_8, G_{10}$ and every nonisomorphic graph on at most seven vertices.

\begin{table}[ht]
\centering
\renewcommand{\arraystretch}{1.15}
\setlength{\tabcolsep}{12pt}
\begin{tabular}{lcc}
\toprule
Solver & Instance & Running time \\
\midrule
Python (reference) & $G_{10}$ & $\sim 2$ min \\
Lean (verification) & $G_{10}$ & $\sim 2$ h \\
C++ (multithreaded) & $B_{2,3,3}$ & $\sim 14$ min (128 threads, 63\,GiB memory) \\
\bottomrule
\end{tabular}
\caption{Running times of the three implementations.}
\label{tab:implementations}
\end{table}

\paragraph{Exhaustive search.}
To find worst-case general instances on $n\le11$ vertices and bipartite instances on $n\le12$, we enumerate nonisomorphic nonempty graphs and first screen them using an $O(2^n m)$ subset dynamic program for \Greedy{}: since $V^\star(G)\ge\E[|\mathrm{\Greedy{}}(G)|]$, only candidates whose \Greedy{} ratio falls below the target threshold require exact evaluation of Equation~\eqref{eq:fixed-graph-bellman}.
This screening greatly reduces the number of graphs requiring exact evaluation, e.g., among the $12\,005\,167$ nonisomorphic nonempty graphs on ten vertices, only $21$ have a \Greedy{} ratio below $0.7$.

\section*{Acknowledgments}
\label{sec:acknowledgments}
We are deeply grateful to Zhihao Gavin Tang, Xiaowei Wu, and Yuhao Zhang for many inspiring discussions and helpful comments.

\section*{AI Methodology}
\label{sec:ai-methodology}
We used GPT-6 Astra throughout the development of this paper:
\begin{itemize}
    \item \textbf{Lower bounds.} For the competitive analysis of \AlgorithmName{}, it enumerated local configurations and organized case-analysis tables used to develop the local greedy comparison, and simplified the polynomial-expansion proofs.
    \item \textbf{Upper bounds.} Following our design for parallel memoized search and formal verification, it developed a multithreaded C++ solver and a Lean~4 solver based on our Python reference program, together with correctness proofs checked by Lean~4.
    \item \textbf{Exposition.} It refined the exposition and figures across the paper.
\end{itemize}

\clearpage
\bibliographystyle{plainnat}
\bibliography{refs}

\clearpage
\appendix
\section{Missing Proofs in Section~\ref{sec:local}}
\label{Appendix:A}

In this section, we prove Lemmas~\ref{lem:vertex-insertion}, \ref{lem:one-ref-deletion} and \ref{lem:shared-ref}.

\begin{proofof}{Lemma~\ref{lem:vertex-insertion}}
Fix the edge order $\sigma$ and which edges between $\{x_1,x_2\}$ and $Y$ are retained. Recall that $C=B_e-x_2$. Before the run on $B_e$ first accepts an edge incident to $x_2$, both runs make identical decisions and agree on the matched status of every vertex in $V(C)$. If $x_2$ remains unmatched at the end of the run on $B_e$, both runs agree on all of $V(C)$ and the equality holds trivially.

Suppose instead that $x_2$ is matched in $B_e$, and let $(x_2,u)$ be the first edge incident to $x_2$ accepted in $B_e$. Immediately after $(x_2,u)$ is accepted, $u$ is the unique vertex in $V(C)$ whose matched status differs between the two runs at this time (matched in $B_e$, free in $C$). Next, we use induction on subsequent arrivals to prove that exactly one vertex in $V(C)$ has a different matched status between the two runs at any later time:
\begin{itemize}[topsep=2pt, itemsep=1pt]
    \item Unless the arriving edge belongs to $C$ and is incident to $u$, the unique differing vertex remains unchanged. Edges incident to $x_2$ are rejected in $B_e$ and absent from $C$; all other such edges see identical endpoint statuses in both runs, so both runs make the same decision.
    \item Suppose an edge $(u,w)\in E(C)$ arrives, where $u$ is the current unique differing vertex. Then $w\ne u$ currently has the same status in both runs. If $w$ is already matched in both runs, both runs reject $(u,w)$. If $w$ is free in both runs, the run where $u$ is free accepts $(u,w)$, while the run where $u$ is already matched rejects $(u,w)$; after this step, $u$ is matched in both runs and $w$ becomes the unique differing vertex in $V(C)$.
\end{itemize}
At the end of the run, $x_2$ is matched in $B_e$, contributing $3$ to $3\N+2\R$, while at most one old vertex in $V(C)$ loses its matched status in $B_e$ relative to $C$. Such a loss has weight $3$ if the vertex is $x_1$, weight $2$ if it is a reference vertex, and weight $0$ otherwise. The new contribution of $x_2$ therefore compensates for this loss, proving $3\N+2\R\ge3\N_1+2\R_1.$
\end{proofof}

\begin{proofof}{Lemma~\ref{lem:one-ref-deletion}}
Let $H=B-z$, and assign independent arrival times from $U[0,1]$ to the edges. Let $\tau$ be the time at which \Greedy{} on $H$ matches $y_i$, with $\tau=1$ if $y_i$ remains unmatched. We first prove that $\mu_i\ge1-\E[\tau]$.

Fix the edges accepted by \Greedy{} on $H$ and their arrival times. If $\tau=1$, the conditional lower bound $1-\tau$ is zero. Otherwise, let $(y_i,w)$ be the edge accepted at time $\tau$. For $t<\tau$, let $p(t)$ denote the probability that \Greedy{} on $B$ has not matched $w$ by time $t$, conditional on the fixed edges accepted by \Greedy{} on $H$ and their arrival times.

Further reveal all arrivals in $B$ up to time $t$. While $w$ remains unmatched in $B$, it is unmatched in both runs. If $z$ is still unmatched in $B$, the two runs agree on the matched status of every vertex in $H$, and $z$ is the only additional unmatched vertex in $B$. If $z$ is already matched in $B$, the comparison in the proof of Lemma~\ref{lem:vertex-insertion} ensures that at most one vertex is unmatched in $B$ but matched in $H$. Let $a=z$ in the former case, and let $a$ denote this unique vertex, if it exists, in the latter case. An edge connecting $w$ to a vertex still unmatched in $H$ cannot arrive at this time, since \Greedy{} on $H$ would then match $w$ before $\tau$. Thus, only $(a,w)$ can match $w$ at this time.

If $(a,w)$ exists and has not arrived, its remaining arrival time is uniform on $(t,1)$. Indeed, when $a\ne z$, its fixed matching time in $H$ is already past; when $a=z$, its arrival time is unrestricted. The edgewise conditioning argument of Lemma~\ref{lem:cond-clocks} therefore applies in either case. Choose $\delta>0$ such that $t+\delta<\tau$ and \Greedy{} on $H$ accepts no edge during $(t,t+\delta]$. Throughout this interval, the matched vertices in $H$ remain unchanged, while the unmatched vertices in $B$ can only disappear. Thus, the only edge that can match $w$ during this interval is the same edge $(a,w)$, when it exists. Its conditional probability of arriving during the interval is $\delta/(1-t)$. Averaging over the arrivals revealed up to time $t$, we obtain $p(t+\delta) \ge p(t)\cdot\left(1-\frac{\delta}{1-t}\right)$, which yields $\frac{p(t+\delta)}{1-t-\delta}
\ge\frac{p(t)}{1-t}.$

Hence, $p(t)/(1-t)$ is nondecreasing between consecutive fixed acceptance times of $H$. At each such time before $\tau$, no edge incident to $w$ arrives with positive conditional probability, so $p$ is continuous there. Therefore, $p(t)/(1-t)$ is nondecreasing throughout $[0,\tau)$. Since $p(0)=1$, we conclude that $p(t)\ge 1-t$.

If $w$ is still unmatched before $\tau$, then $y_i$ is either already matched in $B$ or becomes matched when $(y_i,w)$ arrives. Conditional on the accepted edges of $H$ and their arrival times, $y_i$ is matched in $B$ w.p. at least $1-\tau$. Taking expectations, we obtain $\mu_i\ge\E[1-\tau]=1-\E[\tau]$.

We next bound the contribution in $D$. Fix all arrival times in $H$, as well as the arrival times and the decisions on whether to retain each edge incident to $x_i$ other than $(x_i,y_i)$. If \Greedy{} on $D-(x_i,y_i)$ matches $x_i$, adding $(x_i,y_i)$ either leaves the run unchanged or matches $x_i$ earlier. Thus, $3n+2r\ge3\ge2+\tau$.

Otherwise, this run agrees with \Greedy{} on $H$. If $(x_i,y_i)$ is retained and arrives before $\tau$, it is accepted and contributes $5$. If it arrives after $\tau$, then $y_i$ is already matched and contributes $2$. These disjoint events have probabilities $q\tau$ and $1-\tau$, respectively. Since all other contributions are nonnegative,
\begin{align*}
\E[3n+2r\mid\text{the fixed information}]
&\ge5q\tau+2(1-\tau)=2+(5q-2)\tau
\ge2+\tau,
\end{align*}
where the last inequality uses $5q\ge3$. This calculation also covers $\tau=1$.

Taking expectations and combining the two bounds, we have
\begin{align*}
\E[3n+2r]\geq 2+\E[\tau] \geq 3-\mu_i.\tag*{\qedsymbol}
\end{align*}
\let\qed\relax
\end{proofof}

\begin{proofof}{Lemma~\ref{lem:shared-ref}}
Let $\R_{y_2}$ and $\K_{y_2}$ indicate whether $y_2$ is matched in $M(B_e)$ and $M(B)$, respectively. We first prove
\[
\E[3\N+2\R_{y_2}+2\K_{y_2}]\ge4.
\]

Fix which edges between $\{x_1,x_2\}$ and the non-reference vertices are retained. Consider a uniformly random arrival order of the edges of $B$, these retained edges, and the two reference edges $(x_1,y_2)$ and $(x_2,y_2)$. Both reference edges are included in this order, while each is independently retained w.p. $q$. Run \Greedy{} on $B$ until either $y_2$ is matched or an edge from $\{x_1,x_2\}$ to a currently unmatched vertex of $B$ arrives.

Before this point, every arriving edge incident to $x_1$ or $x_2$ has its endpoint in $B$ already matched. Thus, $x_1,x_2$ remain unmatched and the two runs agree on $B$. If $y_2$ is matched at this point, then $2\R_{y_2}+2\K_{y_2}=4$.

Otherwise, assume the current edge is incident to $x_1$. Conditional on the arrived edges and their order, the remaining order is uniform. Neither reference edge arrived earlier, since either would have caused \Greedy{} to stop. The current edge is therefore retained w.p. at least $q$. In the following analysis, we discard the vertices of $B$ already matched in both runs, since they cannot be matched again.

If the current edge is retained, $x_1$ contributes $3$. If it also matches $y_2$, another $2$ is obtained immediately. If it does not match $y_2$, apply Inequality~\eqref{eq:one-reference-bound} to $x_2,y_2$ and the remaining graph after accepting this edge. Their conditional expected contribution is at least $3-1=2$, since the probability that \Greedy{} on this remaining graph matches $y_2$ is at most $1$. Thus, the conditional expected contribution is at least $5$.

If the current edge is not retained, it must be $(x_1,y_2)$. Apply Inequality~\eqref{eq:one-reference-bound} to $x_2,y_2$, then add $x_1$ using Lemma~\ref{lem:vertex-insertion}, counting only $y_2$ as a reference vertex. The conditional expectation of $3\N+2\R_{y_2}+\K_{y_2}$ is at least $3$, and the additional term $\K_{y_2}$ is nonnegative.

Thus, the conditional expected contribution is at least $5q+3(1-q)=3+2q\ge4$. Averaging proves the claimed bound. Since $\R\ge \R_{y_2}$, $\E[\K_{y_2}]=\mu_2$, and $\E[\K]=\mu_1+\mu_2$, we conclude that
\begin{align*}
\E[3\N+2\R+\K]
&\ge \E[3\N+2\R_{y_2}+2\K_{y_2}]+\mu_1-\mu_2\ge4+\mu_1-\mu_2. \qedhere
\end{align*}
\end{proofof}

\section{A Stronger Guarantee for Triangle-Free Graphs}
\label{app:triangle-free}

In this section, we prove Theorem~\ref{thm:main-tf} by fixing the parameters $a=3/32,b=4/11,p=11/25$.
When a shadow edge has two reference vertices, triangle-freeness rules out the cross edges to these vertices, allowing us to strengthen the local comparison from weights $(3,2,1)$ to $(7,4,3)$.

\subsection{\texorpdfstring{{A Stronger Local Comparison}}{A Stronger Local Comparison}}

Fix $H_b$ and a shadow edge $e=(x_1,x_2)\in T_a$.
Condition further on $B$, as in Section~\ref{sec:cond-clocks}.
If $e$ arrives at time $s<a$, each edge between $\{x_1,x_2\}$ and $Y$ is independently retained with probability
\begin{equation*}
    q = \frac{1-b}{1-s}\ge 1-b = \frac{7}{11}.
\end{equation*}
We use the equivalent formulation with a common uniform edge order $\sigma$.
Throughout this subsection, expectations are conditioned on $H_b$ and $B$, and we write $k=k_e$.

Suppose that $k=2$, with $(x_1,y_1),(x_2,y_2)\in M^*$.
Since the original graph contains the shadow edge $(x_1,x_2)$, the only edges between $\{x_1,x_2\}$ and $\{y_1,y_2\}$ before retention are $(x_1,y_1)$ and $(x_2,y_2)$.
Otherwise, it would violate the triangle-freeness.

Recall that $\mu_i$ denotes the probability that the reference vertex $y_i$ is matched in $M(B)$ for each $i\in \{1,2\}$.
Similar to Lemma~\ref{lem:one-ref-deletion}, we show the following bound with different weights on $n$ and $r$.

\begin{lemma}
\label{lem:tf-single-vertex}
Under the setup of Lemma~\ref{lem:one-ref-deletion}, if $q\ge 7/11$, then
\begin{equation*}
    \E[7n+4r]\ge 7-3\mu_i.
\end{equation*}
The bound also holds after fixing which edges incident to $x_i$, other than $(x_i,y_i)$, are retained.
\end{lemma}
\begin{proof}
The proof of Lemma~\ref{lem:one-ref-deletion} applies with weights $7$ and $4$ in place of $3$ and $2$.
The condition that $5q\ge 3$ used there is replaced by $11q\ge 7$, which holds by assumption.
The same argument therefore proves the statement.
\end{proof}

\begin{lemma}
\label{lem:tf-pair-comparison}
For the coupled matchings $M(B)$ and $M(B_e)$, we have
\begin{equation*}
    \E[7\N+4\R+3\K]\ge 7k.
\end{equation*}
\end{lemma}

\begin{proof}
The case when $k=0$ is immediate.

For $k=1$, use the notation and the three cases from Section~\ref{sec:one-ref}, with weights $7,4,3$ on $\N_1,\R_1,\K$, respectively.
In the first two cases, the contribution is at least $7$.
In the third case, its conditional expectation is
$11q\ge 7$.
Thus, we have $\E[7\N_1+4\R_1+3\K]\ge 7$.
The proof of Lemma~\ref{lem:vertex-insertion} also applies
with endpoint weight $7$ and reference-vertex weight $4$: the added endpoint contributes $7$, while at most one old vertex loses its matched status, with weight at most $7$.
Hence, adding $x_2$ gives $7\N+4\R\ge 7\N_1+4\R_1$, which proves the claim for $k=1$.

Now suppose that $k=2$.
Recall the definition of $L$ from Section~\ref{sec:two-ref}.
Fix the retention choices for edges from $\{x_1,x_2\}$ to $Y\setminus\{y_1,y_2\}$.
For a triangle-free graph, the only counted edges between $\{x_1,x_2\}$ and $\{y_1,y_2\}$ are $(x_1,y_1)$ and $(x_2,y_2)$.
As in Section~\ref{sec:two-ref}, we prove the lemma when $k=2$ by induction on $L$.
When $L=2$, each reference edge contributes $11$ whenever retained, so
\begin{equation*}
    \E[7\N+4\R+3\K]=22q\ge 14.
\end{equation*}
For $L>2$, condition on the first of the $L$ counted edges in $\sigma$.
When the first edge is incident to $x_1$ or $x_2$, the relative order within $B$ remains uniform.
Thus, conditional on this first edge, the probability that $M(B)$ matches $y_i$ is still $\mu_i$ for each $i\in\{1,2\}$.

If the first edge belongs to $B$, the same argument as in Section~\ref{sec:two-ref}, with weights $7,4,3$, gives a conditional expected contribution of at least $14$.
If the first edge is $(x_1,z)$ with $z\in Y\setminus\{y_1,y_2\}$, Lemma~\ref{lem:tf-single-vertex} and using $\E[\K]=\mu_1+\mu_2$ give a lower bound of
\begin{equation*}
    7+(7-3\mu_2)+3(\mu_1+\mu_2) = 14+3\mu_1\ge 14.
\end{equation*}

It remains to consider the case when the first edge joins an endpoint of $e$ to a reference vertex.
By symmetry, let it be $(x_1,y_1)$.
If retained, it is accepted and contributes $11$ from $x_1,y_1$ in $M(B_e)$.
\Greedy{} continues on $B-y_1$ together with $x_2$ and its retained edges.
Applying Lemma~\ref{lem:tf-single-vertex} with $i=2$ and $z=y_1$ gives a lower bound of
\begin{equation*}
    11+(7-3\mu_2)+3(\mu_1+\mu_2) = 18+3\mu_1.
\end{equation*}
If $(x_1,y_1)$ is not retained, apply the $k=1$ case with only $y_2$ counted as a reference vertex.
The two endpoints, $y_2$, and its baseline contribution give at least $7$ in expectation.
The baseline contribution of $y_1$ adds $3\mu_1$, while its contribution in $M(B_e)$ is nonnegative.
Taking the expectation over whether $(x_1,y_1)$ is retained gives
\begin{equation*}
    q\cdot (18+3\mu_1)+(1-q)\cdot (7+3\mu_1) = 7+11q+3\mu_1\ge 14.
\end{equation*}

Therefore, every possible first edge has conditional expected contribution at least $14$, completing the induction.
Finally, taking expectations over the retention choices for edges to non-reference vertices proves the lemma.
\end{proof}

\subsection{The Global Bound}

We now take the expectation over $B$ and over the reservation choice of $e$.
As in Section~\ref{sec:pair-recovery},
\begin{equation*}
    7\cdot \E[\hat \N_e\mid H_b] + 4\cdot \E[\hat \R_e\mid H_b] =
    p\cdot \E[7\N+4\R+3\K\mid H_b] + (4-7p)\cdot \E[\K\mid H_b] \ge \frac{77}{25}\cdot k_e.
\end{equation*}

Recall that $\R_{1Y} = \sum_{e\in T_a}\hat \R_e$.
Summing over $e\in T_a$ gives
\begin{equation*}
    7\cdot \E[Z_{XY}\mid H_b] + 4\cdot \E[\R_{1Y}\mid H_b] \ge \frac{77}{25}\cdot |M^*_{1Y}|.
\end{equation*}
The vertex count from Section~\ref{sec:suffix-agg} gives
$\R_{1Y}+|M^*_{YY}|\le Z_{XY}+2Z_{YY}$.
Consequently,
\begin{equation*}
    \frac{77}{25}\cdot |M^*_{1Y}| + 4\cdot |M^*_{YY}|
    \le 11\cdot \E[Z_{XY}\mid H_b] + 8\cdot \E[Z_{YY}\mid H_b]
    \le 11\cdot \E[Z_{XY}+Z_{YY}\mid H_b].
\end{equation*}

Combining the above inequality with Equation~\eqref{eq:prefix-suffix-decomposition}, Lemma~\ref{lem:count1}, and Lemma~\ref{lem:count2}, we have
\begin{align*}
    \E[\ALG\mid H_b] &\ge |T_b|-\frac{11}{25}\cdot |T_a| + \frac{7}{25}\cdot |M^*_{1Y}| + \frac{4}{11}\cdot U_b \\
    &\ge \frac{14}{25}\cdot \OPT + \frac{11}{25}\cdot (|T_b|-|T_a|) - \frac{7}{25}\cdot U_a + \frac{7}{25}\cdot |M^*_2| + \frac{23}{275}\cdot U_b \\
    &\ge \frac{14}{25}\cdot \OPT + \frac{1}{2}\cdot |M^*_2|-\frac{3}{50}\cdot U_a - \frac{3}{22}\cdot U_b.
\end{align*}
Taking expectations over $H_b$ and applying Equations~\eqref{eq:expected-U} and~\eqref{eq:expected-M2}
therefore gives
\begin{equation}
\label{eq:tf-global-bound}
    \frac{\E[\ALG]}{\OPT} \ge \frac{14}{25} + \frac{1}{2}\int_a^b \Survival(t)\,dt - \frac{3}{50}(1-a)\cdot \Survival(a)
    -\frac{3}{22}(1-b)\cdot \Survival(b).
\end{equation}

\subsection{Verifying the Guarantee}

Recall that we fix $a = 3/32$ and $b = 4/11$.
For every integer $n\ge 1$, define
\begin{equation*}
    L_n^{\mathrm{tf}} = \frac{14}{25} + \frac{(1-a)^n-(1-b)^n}{2n} - \frac{3}{50}(1-a)^n - \frac{3}{22}(1-b)^n.
\end{equation*}
Substituting Equation~\eqref{eq:r-power-expansion} into Inequality~\eqref{eq:tf-global-bound}, we obtain
\begin{equation*}
    \frac{\E[\ALG]}{\OPT} \ge \sum_{n\ge 1} c_n\cdot L_n^{\mathrm{tf}}, \qquad
    c_n\ge 0,\qquad \sum_{n\ge 1} c_n=1.
\end{equation*}
It remains to show that $L_n^{\mathrm{tf}}>0.5535$
for every integer $n\ge1$.

For $1\le n\le 14$, direct computation gives the following lower bounds on $L_n^{\mathrm{tf}}$, all of which exceed $0.5535$.
\begin{center}
\renewcommand{\arraystretch}{1.15}
\setlength{\tabcolsep}{12pt}
\begin{tabular}{cc@{\qquad}cc}
\toprule
$n$ & Lower bound & $n$ & Lower bound \\
\midrule
1 & 0.553791 & 8  & 0.555789 \\
2 & 0.559583 & 9  & 0.554882 \\
3 & 0.561300 & 10 & 0.554233 \\
4 & 0.560982 & 11 & 0.553814 \\
5 & 0.559784 & 12 & 0.553588 \\
6 & 0.558335 & 13 & 0.553519 \\
7 & 0.556956 & 14 & 0.553571 \\
\bottomrule
\end{tabular}
\end{center}

For $n\ge 14$, write
\begin{equation*}
    L_n^{\mathrm{tf}} = \frac{14}{25}-A_n-B_n, \qquad
    A_n = \left(\frac{3}{50}-\frac{1}{2n}\right)\cdot (1-a)^n, \qquad
    B_n = \left(\frac{3}{22}+\frac{1}{2n}\right)\cdot (1-b)^n.
\end{equation*}
The sequence $B_n$ is decreasing.
Moreover,
\begin{equation*}
    A_{n+1}\le A_n \quad \Longleftrightarrow \quad \frac{3n}{32}\cdot (3n-22)\ge 25.
\end{equation*}
At $n=14$, the LHS is $105/4>25$, and it increases with $n$ for $n\ge 14$.
Thus, $A_n$ is also decreasing for $n\ge 14$.
It follows that $L_n^{\mathrm{tf}}$ is increasing for $n\ge14$.

We have therefore proved $L_n^{\mathrm{tf}}>0.5535$ for every integer $n\ge1$.
Since the coefficients $c_n$ are nonnegative and sum to $1$, Theorem~\ref{thm:main-tf} follows.

\end{document}